\documentclass[]{spie}  %>>> use for US letter paper
\usepackage{amsmath,amsfonts,amssymb}
\usepackage{graphicx}
\usepackage{color,soul}
\usepackage[colorlinks=true, allcolors=blue]{hyperref}
\usepackage{tikz}
\usepackage[nolist]{acronym}
\usepackage[acronym]{glossaries}
\usepackage{url}
\usepackage{multirow}
\usepackage{color, colortbl}
\usepackage{subfig}
\usepackage{relsize}
\usepackage[pagewise]{lineno}
\newacro{pywfs}[PyWFS]{pyramid wavefront sensor} 
\newacro{dm}[DM]{deformable mirror} 
\newacro{ao}[AO]{adaptive optics}
\newacro{exao}[ExAO]{extreme adaptive optics}
\newacro{scexao}[SCExAO]{Subaru Coronagraphic Extreme Adaptive Optics system}
\newacro{mvm}[MVM]{matrix vector multiplication}
\newacro{rmse}[RMSE]{root mean squared error}
\newacro{nn}[NN]{neural network}
\newacro{nns}[NNs]{neural networks}
\newacro{rnn}[RNN]{recurrent neural network}
\newacro{cnn}[CNN]{convolutional neural network}
\newacro{shwfs}[SHWFS]{Shack-Hartmann wavefront sensor}
\newacro{rms}[RMS]{root mean squared}
\newacro{wfs}[WFS]{wavefront sensor}
\newacro{magao}[MagAO]{Magellan Adaptive Optics system}
\newacro{lbtao}[LBTAO]{Large Binocular Telescope Adaptive Optics system}
\newacro{gmt}[GMT]{Giant Magellan Telescope}
\newacro{elt}[ELT]{European Extremely Large Telescope}
\newacro{tmt}[TMT]{Thirty Meter Telescope}
\newacro{lips}[LIPS]{Landweber Iteration for Pyramid Sensors}
\newacro{klips}[KLIPS]{Kaczmarz-Landweber Iteration for Pyramid Sensors}
\newacro{dl}[DL]{deep learning}
\newacro{bfgs}[BFGS]{Broyden-Fletcher-Goldfar-Shanno}
\newacro{mmt}[MMT]{Multiple Mirror Telescope}
\newacro{gpu}[GPU]{Graphics Processing Unit}
\newacro{svd}[SVD]{singular value decomposition}

\title{Nonlinear wavefront reconstruction from a pyramid sensor using neural networks}

\author[a, b]{Alison P. Wong}
\author[a, b, c]{Barnaby R. M. Norris}
\author[d]{Vincent Deo}
\author[a]{Peter G. Tuthill}
\author[e]{\\Richard Scalzo}
\author[a, b]{David Sweeney}
\author[d]{Kyohoon Ahn}
\author[d]{Julien Lozi}
\author[d, f]{\\S\'ebastien Vievard}
\author[d, f, g, h]{Olivier Guyon}

\affil[a]{Sydney Institute for Astronomy, School of Physics, Physics Road, University of Sydney, NSW 2006, Australia}
\affil[b]{Sydney Astrophotonic Instrumentation Laboratories, Physics Road, University of Sydney, NSW 2006, Australia}
\affil[c]{Astralis, School of Physics, University of Sydney 2006}
\affil[d]{National Astronomical Observatory of Japan, National Institutes of Natural Sciences, Subaru Telescope, 650 North A'oh\=ok\=u Place, Hilo, HI 96720, U.S.A.}
\affil[e]{Centre for Translational Data Science, University of Sydney, Darlington NSW 2008, Australia}
\affil[f]{Astrobiology Center of NINS, 2-21-1, Osawa, Mitaka, Tokyo, 181-8588, Japan}
\affil[g]{College of Optical Sciences, University of Arizona, Tucson, AZ 85721, U.S.A}
\affil[h]{Steward Observatory, University of Arizona, Tucson, AZ 85721, U.S.A}

\authorinfo{Further author information: (Send correspondence to A.P.W.)\\A.P.W.: E-mail: a.wong@sydney.edu.au, Telephone: +61 414 211 546}

\begin{document} 

\maketitle

\begin{abstract}

The \ac{pywfs} has become increasingly popular to use in \ac{ao} systems due to its high sensitivity. The main drawback of the \ac{pywfs} is that it is inherently nonlinear, which means that classic linear wavefront reconstruction techniques face a significant reduction in performance at high wavefront errors, particularly when the pyramid is unmodulated. In this paper, we consider the potential use of \ac{nns} to replace the widely used \ac{mvm} control. We aim to test the hypothesis that the \ac{nn}'s ability to model nonlinearities will give it a distinct advantage over \ac{mvm} control. We compare the performance of a \ac{mvm} linear reconstructor against a dense \ac{nn}, using daytime data acquired on the \ac{scexao} instrument. In a first set of experiments, we produce wavefronts generated from 14 Zernike modes and the \ac{pywfs} responses at different modulation radii (25, 50, 75, and 100 mas). We find that the \ac{nn} allows for a far more precise wavefront reconstruction at all modulations, with differences in performance increasing in the regime where the \ac{pywfs} nonlinearity becomes significant. In a second set of experiments, we generate a dataset of atmosphere-like wavefronts, and confirm that the \ac{nn} outperforms the linear reconstructor. The \ac{scexao} real-time computer software is used as baseline for the latter. These results suggest that \ac{nn}s are well positioned to improve upon linear reconstructors and stand to bring about a leap forward in \ac{ao} performance in the near future.

\end{abstract}
% focus on interpretation
% bench data


% feats

% --------------------
\section{Introduction}
% --------------------

\ac{ao} systems are essential to the scientific productivity of many ground-based optical/IR telescopes, as they enable diffraction-limited imaging performance despite the angular smearing caused by seeing, which would otherwise limit resolution to approximately 0.5-2 arcseconds. The main components that make up an \ac{ao} system are: a \ac{wfs} that measures instantaneous phase aberrations, a \ac{dm} which rapidly corrects aberrations upon reflection, and a control system that determines the \ac{dm} shape from the wavefront sensor. The post-\ac{ao} image quality relies on the accuracy to which aberrated wavefronts can be measured, reconstructed, and compensated; it is therefore ideal to employ highly sensitive wavefront sensors and effective reconstruction techniques.

Employment of \ac{pywfs}s\cite{ragazzoni1996} has increased as \ac{ao} systems have begun to trade the more conventionally used \ac{shwfs}\cite{Platt2001HistoryAP, guyon2005, 2004SPIE.5490.1177V, 2000SPIE.4007..416E} for the \ac{pywfs}, which offers greater sensitivity. The \ac{pywfs} also offers users flexibility as its sensitivity and dynamic range can be tuned through modulation and, when used with a charged-coupled device (CCD) camera, the resolution of the \ac{pywfs} image can be changed to accommodate various noise levels. \ac{pywfs}s are in use at some of the world's leading general and high-contrast \ac{ao} facilities\cite{Esposito2011, Close2013, 2021SPIE11823E..02W}, including  \ac{scexao}\cite{Jovanovic2015}.
\ac{pywfs}s are also the \ac{wfs} of choice for most instruments of upcoming extremely large telescopes such as the \ac{gmt}\cite{Esposito2012, 2018SPIE10703E..0WB}, the \ac{tmt}\cite{boyer2018, 2018SPIE10703E..3VC} and the \ac{elt}\cite{hadi2013, 2021Msngr.182...17D, 2021Msngr.182...22B}.
These telescopes are $\sim4$ times larger in diameter than the current largest optical telescopes and will benefit greatly from the improved resolution enabled by the \ac{ao} systems of tomorrow. The advent of the ELTs with their AO systems will enable diffraction-limited imaging with visible-wavelength angular resolutions better than $\sim$10~mas.

One major challenge that comes with using a \ac{pywfs} is that it is an inherently nonlinear sensor (approximately linear only for small wavefront aberrations) while most of the commonly used wavefront reconstruction techniques are linear.
Modulation -- steering the beam using a rapid scanning tip-tilt mirror synchronised with the acquisition camera -- increases the linear regime of the \ac{pywfs}, but at the cost of sensitivity. The modulation amplitude can be chosen based on seeing conditions to maximize sensitivity while maintaining favorable conditions for stable closed-loop control with a linear reconstructor. No facility PyWFS AO system, to date, has achieved a consistent, reliable operational performance while using a PyWFS at its maximum, unmodulated sensitivity. Yet, there is strong interest in achieving such a design, particularly to meet the requirements of \ac{exao} systems. 

This has motivated the development of various nonlinear wavefront reconstruction techniques\cite{Hutterer2018Nonlinear}. Iterative phase retrieval algorithms\cite{Gerchberg1972A} are not amenable to fast real-time use, and approaches have been proposed to address this via pre-computation of optical systems and solving with quasi-Newton methods\cite{Frazin2018Phase}. Another approach consists in dynamically adapting a linear reconstruction law to accommodate for nonlinear changes in the \ac{pywfs} first-order response\cite{2021A&A...650A..41D, 2020A&A...644A...6C}. Predictive control algorithms\cite{2017arXiv170700570G, 2021SPIE11823E..06H} can also help tackle the nonlinearity problem.

Reconstruction techniques have more recently turned to applying methods that have emerged from \ac{dl}. Early examples where \ac{dl} has been applied to \ac{ao} include the design of \ac{nn}s to predict Zernike coefficients from in-focus and out-of focus images \cite{angel1990, sandler1991, lloydhart1992} or off-axis \ac{wfs} measurements \cite{osborn2014}.
Dense \ac{nn}s have also been applied to wavefront reconstruction from \ac{shwfs}s\cite{guo2006, xu2019, xu2020}, along with newer approaches combining dense \ac{nn}s with reinforcement learning\cite{Nousiainen:21, nousiainen2022, Pou2022Adaptive}.
More recently, \ac{nn}s with more complex architectures have been applied to wavefront prediction and reconstruction. This includes the use of \ac{rnn}s and \ac{cnn}s on e.g. \ac{shwfs}\cite{swanson2018,He:21,Escobar2021Wavefront} data; the PSF\cite{Nishizaki:19}; or pairs of in-focus and out-of-focus images\cite{xin2019}.
Work is also being done on the recovery wavefronts of from a single in-focus PSF\cite{Paine:18, guo2019, andersen2019}.

Previous research has already demonstrated wavefront reconstruction from \ac{pywfs} measurements using \ac{nn}s, first using \ac{cnn}s to recover wavefronts from a three-sided \ac{pywfs}\cite{Alvarez_Diez_2008} for opthalmologic applications. More recently, a \ac{cnn} has been used in combination with an \ac{mvm} prediction and has shown that a model combining both improves the reconstruction for large wavefront aberrations where the \ac{pywfs} response is nonlinear\cite{landman2020} in closed-loop operation.  Additionally, a comparative study was conducted that investigated the performance of a number of \ac{nn} architectures in their ability to reconstruct wavefronts in the Zernike basis from both a \ac{shwfs} and \ac{pywfs}\cite{Escobar2021Wavefront}.

In a tangentially related and rapidly growing area of research, automatic differentiation has proven to be a powerful tool for the optimisation of optical systems\cite{pope2021,Wong2021Phase}. This has since been used to jointly optimize the sensitivity of Fourier-filtering wavefront sensors and their reconstructor and are able to achieve greater sensitivity compared to the \ac{pywfs}\cite{Landman2022}. Reinforcement learning is also surfacing as a promising tool in \ac{ao} control, which could be used in conjunction with a \ac{pywfs}\cite{Landman2021Self}.

In this paper we compare the performance of various \ac{nn}s against linear reconstructors, using data obtained using the \ac{scexao} system at the Subaru Telescope, in off-sky testbed mode. Test wavefronts were applied to the system's \ac{dm} and measured using the system's \ac{pywfs} for a range of modulation amplitudes. We then measured the accuracy to which these wavefronts could be reconstructed.

In Section~\ref{sec:low-order-expts} we study the reconstruction of low-order wavefront errors comprised of Zernike polynomials and compare the reconstruction accuracy between methods and for different pyramid modulation amplitudes. We also compare the performance of a NN predicting modal coefficients, and one directly predicting actuator values. In Section~\ref{sec:seeing-expts} we repeat this analysis using realistic, high-order wavefront error data created by applying von Kármán phase screens to the SCExAO DM. We show that the \ac{nn} greatly out-performs linear reconstructors at low modulation amplitudes, and at high wavefront error. 

% --------------------
\section{Methodology}\label{sec:methodology}
% --------------------

The data used for our experiments were obtained using the \ac{scexao} system at the Subaru Telescope. Using this system, we acquired data consisting of wavefronts applied by the system's \ac{dm}, together with corresponding \ac{pywfs} signals recorded at a wavelength of 750\,nm and a 50\,nm bandwidth. 

SCExAO is equipped with a 2040-actuator Boston Micromachines \ac{dm}, with a $50\times50$ square pitch grid layout. Of these, 1365 actuators lie within the illuminated area of the pupil and are considered active. The \ac{rms} reconstruction errors and the wavefront \ac{rms}s are calculated using only these active elements.
The \ac{pywfs} modulates the beam via a 2-axis piezo steering mirror and uses an OCAM$^2$K EMCCD camera from First Light Imaging as its detector. The frame rate (and modulation speed) used was 1\,kHz. For further details on the SCExAO system, see Ref.~\citenum{2021SPIE11823E..03A}.

\begin{figure}
	\centering
	\includegraphics[width=14cm]{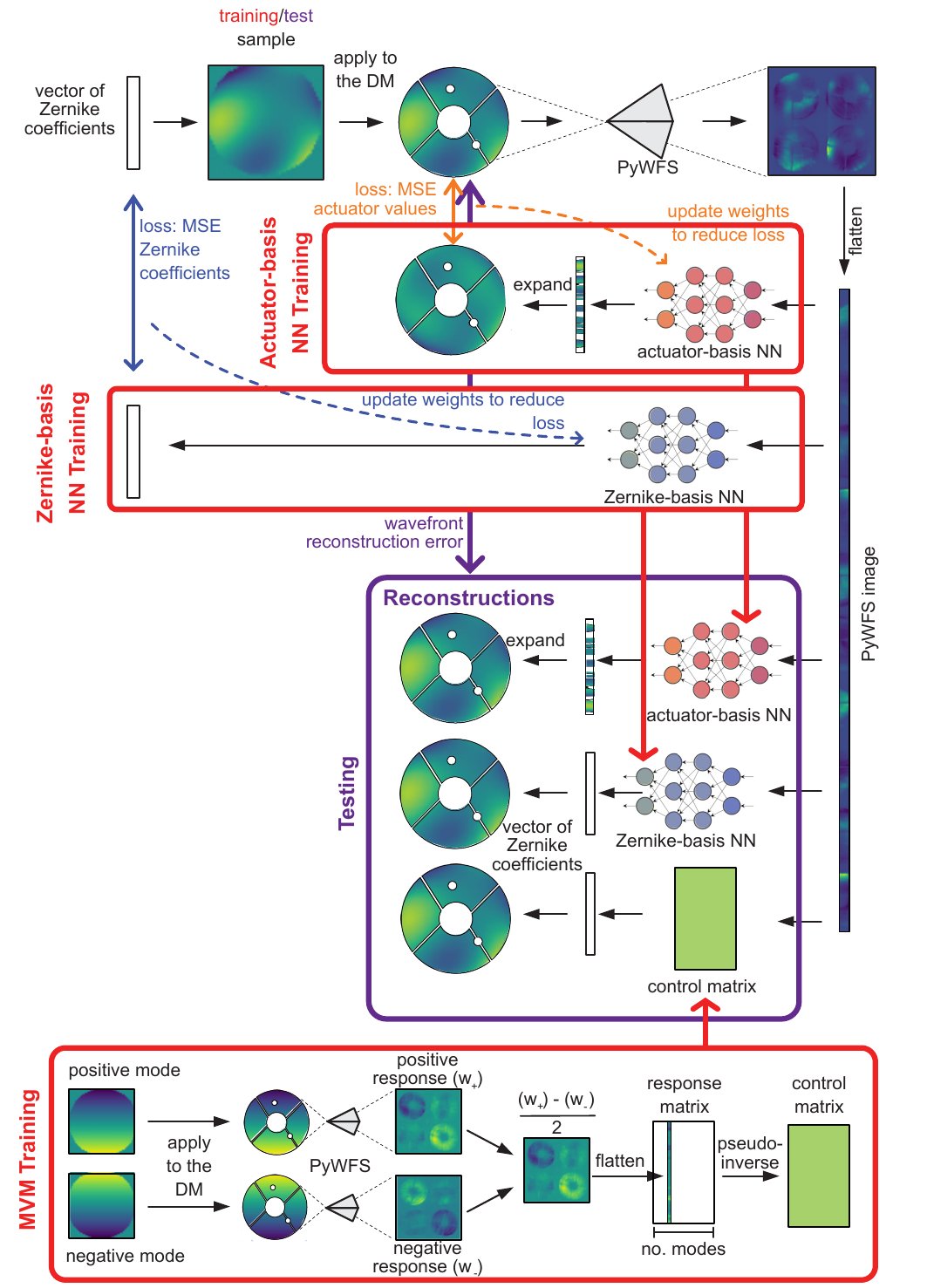}
	\caption{Visualisation of the training and testing process on the low order Zernike wavefronts as described in Sec.~\ref{sec:methodology} and Sec.~\ref{sec:low-order-expts}. For each dataset, we built 3 models: the MVM, Zernike-basis NN and actuator-basis NN.  The MVM was `trained' by calculating a response matrix as described in Sec.~\ref{sec:linear_reconstructors}. The NN models were trained via backpropagation on a dataset consisting of 70~000 pairs of wavefronts and their corresponding \ac{pywfs} measurements. The Zernike-basis NN was trained on wavefronts represented by Zernike mode coefficients while the actuator-basis NN was trained using the pixel values of the wavefronts. At test time, all three models were used to reconstruct wavefronts from \ac{pywfs} images, which were then compared to the true wavefronts.  The results of these models for each modulation are shown in Figure~\ref{fig:zernike_reconstruction}.}
	\label{fig:block1}
\end{figure}

\begin{figure}
	\centering
	\includegraphics[width=14cm]{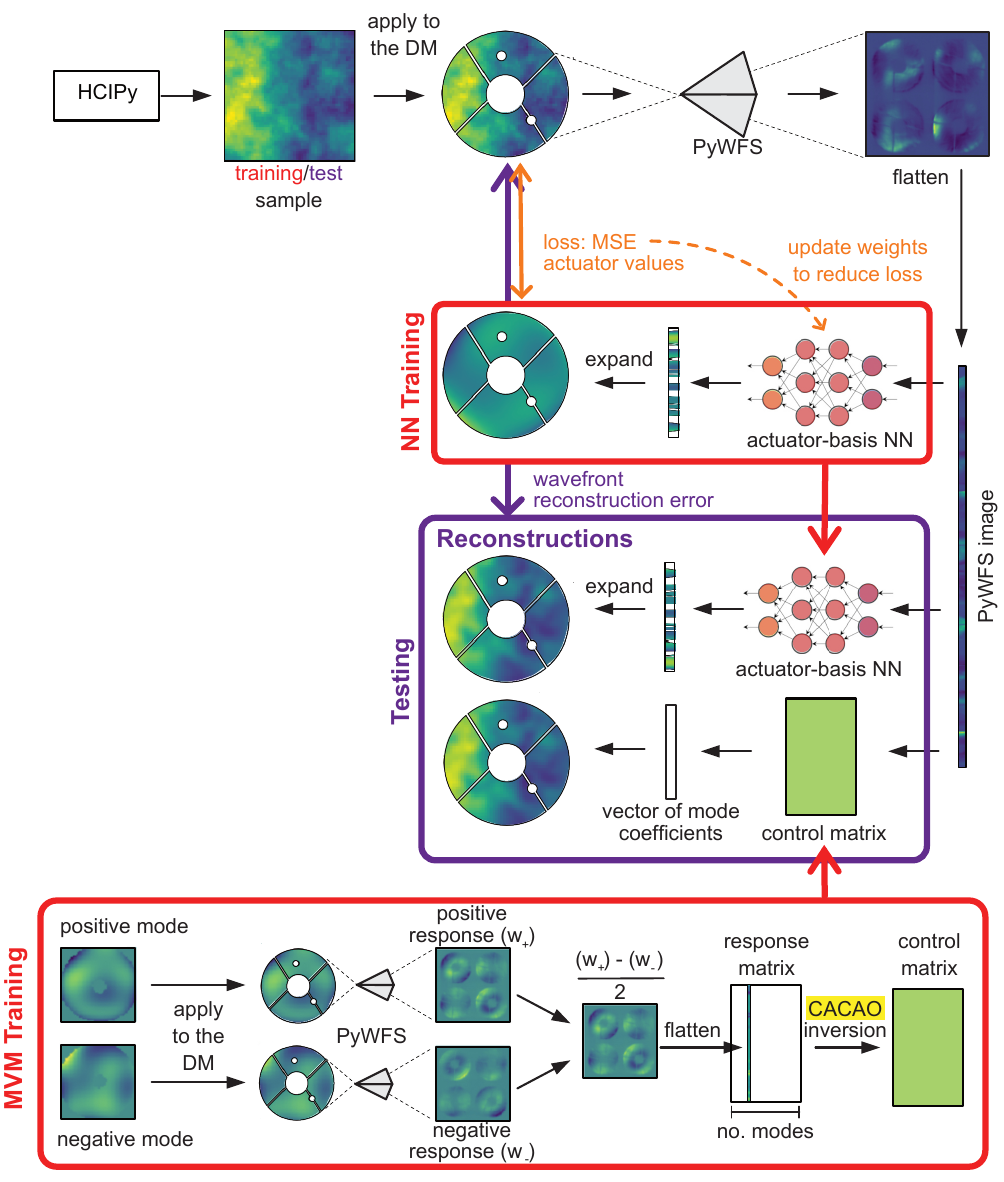}
	\caption{Visualisation of the training and testing process on the turbulence-like dataset as described in Sec.~\ref{sec:methodology} and Sec.~\ref{sec:seeing-expts}. For each dataset, we built 2 models: the MVM and actuator-basis NN. The MVM was `trained' by calculating a response matrix as described in Sec.~\ref{sec:linear_reconstructors}. The NN was trained via backpropagation on a dataset consisting of 170~000 pairs of wavefronts and their corresponding \ac{pywfs} measurements.  At test time, both models were used to reconstruct wavefronts from \ac{pywfs} images, which were then compared to the true wavefronts.  The results of these models for each modulation are shown in Figures~\ref{fig:l020_heatmaps} and \ref{fig:l01_heatmaps}.}
	\label{fig:block2}
\end{figure}

The \ac{scexao} bench was used to produce two kinds of wavefront condition: a low-order Zernike dataset and a dataset intended to mimic atmospheric turbulence. For each dataset, we built linear reconstructor models (described in Sec.~\ref{sec:linear_reconstructors}) and \ac{nn}s (described in Sec.~\ref{sec:NNImpl}). For clarity, we provide visualisations of the training and testing processes in Figures~\ref{fig:block1} and \ref{fig:block2}.

\subsection{Linear reconstructors}\label{sec:linear_reconstructors}
\begin{figure}
	\centering
	\includegraphics[width=17cm]{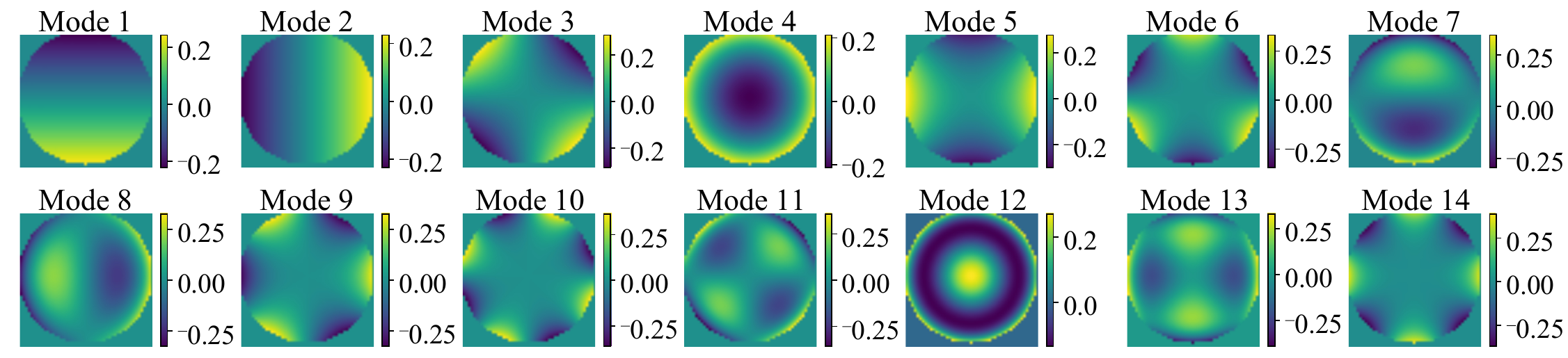}
	\caption{The first 14 Zernike modes (excluding piston) as used on the \ac{scexao} 50 $\times$ 50 actuator grid.
		The color scale is in units of microns.
		Each mode is normalised to 1 radian wavefront \ac{rms} at $\lambda$ = 750~nm.}
	\label{fig:zernike_modes}
\end{figure}

For each of the experiments described in this paper, we produced a simulated dataset using the \ac{scexao} bench as well as measuring a conventional response matrix for the linear reconstructor. The dataset was used to train the \ac{nn}, and to evaluate the success of both the \ac{nn} and linear methods. This allows us to benchmark the \ac{nn} performance against linear reconstructors. 
For the low-order Zernike set, a response matrix was acquired by a classical push-pull method around the best-effort flat wavefront used as reference on SCExAO (which provides $\gtrapprox97$\% H-band Strehl). The 14 first non-piston Zernike terms were probed, as shown on Fig.~\ref{fig:zernike_modes}. These also formed the basis for wavefront generation for our dataset.
The system control matrix, which lets us retrieve \ac{dm}-space wavefront maps from normalised PyWFS measurements, was obtained by direct inversion of the response matrix. The restriction to 14 modes against a high-order PyWFS guaranteed proper conditioning of the matrix.

 % Zernike modes

\begin{figure}
	\centering
	\includegraphics[width=17cm]{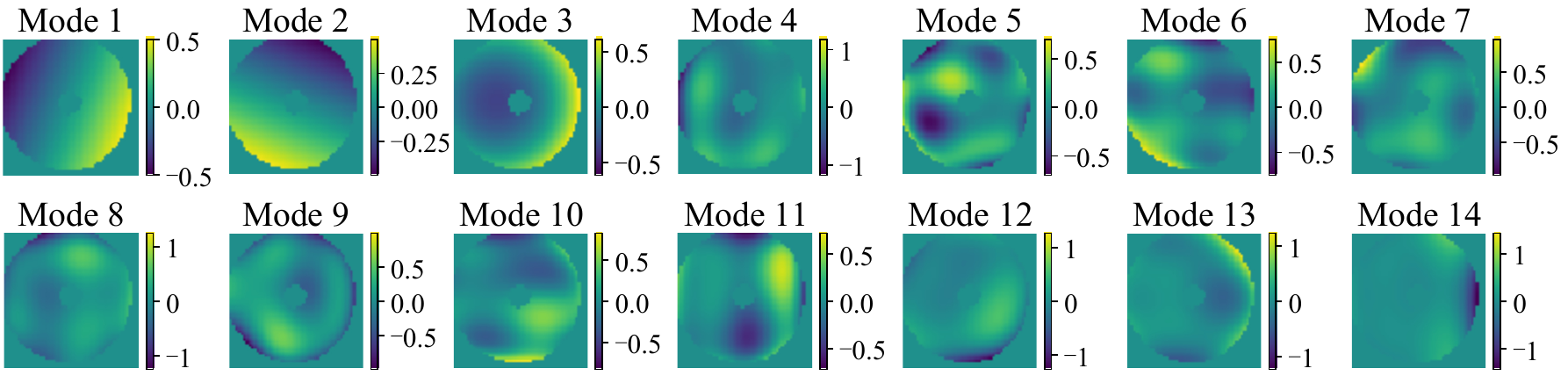}
	\caption{The first 14 modes obtained through the CACAO response matrix calibration method, which is used as a benchmark for the neural network method. The color scale is in units of microns. These modes are normalised to 1 rad \ac{rms} wavefront at $\lambda$ = 750~nm.}
	\label{fig:mvm_modes}
\end{figure}

For the turbulence-like dataset, we used the existing \ac{ao} control software used for \ac{scexao} operations, the Compute-and-control for Adaptive Optics (CACAO\cite{2018SPIE10703E..1EG}) software. 
A more complete description of the \ac{ao} calibration process using CACAO can be found in Ref.~\citenum{2018SPIE10703E..1EG}. A response matrix is measured from push-pull sequences of a basis of Hadamard modes; the control matrix is obtained through a block-wise double diagonalisation of the response matrix, applying \ac{svd} truncations on each of 15 blocks spanning $\sim$1500 modes. 
For illustrative purposes, we plot in Fig.~\ref{fig:mvm_modes} the 14 lowest order modes obtained through this double-diagonalisation inversion process.

\subsection{Neural network implementations}
\label{sec:NNImpl}
For comparison with the \ac{mvm} linear reconstructors, we implemented and trained dense \ac{nn}s. Two models were evaluated --- one which predicted Zernike mode coefficients, and one which directly predicted the 50$\times$50 \ac{dm} actuator map. Our \ac{nn}s followed a simple architecture and comprised of fully connected layers of neurons. The input to each network was 14~400 flattened pixel values from a 120$\times$120 \ac{pywfs} image and the output was either the set of mode coefficients or the set of 2500 DM actuator values.

Reconstruction in the Zernike mode basis was performed for the low order experiments using the same basis that was used to generate the wavefronts (\emph{low-order Zernike-basis NN} hereafter). This is advantageous for basic testing as it provided a complete reconstruction, with no loss of modal subspace between the introduced modes and the reconstructed modes. We later switched to perform wavefront reconstruction in the actuator basis where the network directly outputs stroke commands for each actuator (\emph{low-order actuator-basis NN}). This is more realistic, because typically the optimal mode basis is unknown. Additionally, the actuator basis is complete and can be easily extended to real seeing data, which we demonstrate in Section~\ref{sec:seeing-expts}. 

The number of hidden layers were adjusted per experiment and their exact architectures are detailed in the respective sections of this paper (Section~\ref{sec:low-order-expts} and Section~\ref{sec:seeing-expts}). The \ac{nn}s were implemented with a ReLU\cite{Nair2010Rectified} activation function, which gave them the advantage over the linear reconstructors as it allowed the \ac{nn}s to model nonlinearities. To improve model performance, batch normalisation \cite{Ioffe2015Batch} was used between the hidden layers. All networks were implemented using Keras\cite{Keras2015}, with the Tensorflow\cite{Tensorflow2015} backend. Training and predictions were run on an Nvidia 2080Ti \ac{gpu}.

A key advantage of \ac{nn}s over linear reconstructors is that they have much more powerful regularisation capabilities \cite{Wong2021}. Regularisation of the \ac{mvm} models was achieved by truncating singular values in the \ac{svd} computation as the small singular values tend to encode noise. \ac{nn}s can represent more complex relationships, and so regularisation is key to good performance by preventing overfitting. Methods of regularisation used in regression L1 (lasso)\cite{Tibshirani1996Lasso}, L2 (ridge) or a combination (elastic net) are often applied to \ac{nn}s. L1 regularisation penalises the norm of the weights in the \ac{nn}, which encourages sparsity and L2 regularisation penalise the square of the weights in the \ac{nn}, which encourages edges in the \ac{nn} graph to be close to 0. Elastic net is a weighted combination of the L1 and L2 regularisation. A more popular method of \ac{nn} regularisation is dropout \cite{srivastava2014a}, in which neurons (or nodes) in the \ac{nn} are randomly removed i.e., `dropped out' during training so that the \ac{nn} doesn't become heavily reliant over a subset of the neurons, and effectively `trains' sub-networks within the \ac{nn}. As a result, the \ac{nn} behaves like an ensemble of smaller networks at test time. To determine the optimal method of regularisation and for hyperparameter tuning we performed holdout validation. This is where we put aside a small portion of the data during model training so that we could evaluate our models on unseen data and select the model with the best generalisation capabilities. We found that in all experiments L1 regularisation was the most effective. 

\ac{nn}s work well in practice, but often with little insight into how or why. To gain an understanding of what a \ac{nn} might learn from our data, we also implemented a \emph{bottleneck-\ac{nn}}. This is a \ac{nn} where the network has its smallest layer (least number of neurons) at the middle of the network, hence creating a `bottleneck' through which information passes through. To gain insight into how a \ac{nn} `thinks' the bottleneck can be examined to reveal learned representations of the network, because the bottleneck forces the \ac{nn} to learn an efficient representation of the information it carries. This network, and its results, are described in Section \ref{sec:bottleneck}.

% --------------------
\section{Reconstruction of low order Zernike wavefronts}\label{sec:low-order-expts}
% --------------------

\subsection{Dataset}

100~000 wavefronts were generated from a linear combination of the first 14 Zernike modes where the coefficient, $c$, for each mode was generated from a uniform random distribution ($c\sim\mathcal{U}[-1, 1]$).
Wavefront maps were then scaled such that the wavefront \ac{rms} of the samples was approximately uniformly distributed between 0 and 3.25 radians. These samples were broken up into 70~000 training samples, 10~000 validation samples and 20~000 test samples. 

Data acquisition was performed with the \ac{pywfs} at four modulation radii: $25$, $50$, $75$ and $100$ mas (typical on-sky modulation at \ac{scexao} is $75$ mas), which corresponds to 1.29, 2.59, 3.87 and 5.17 $\lambda/D$ for $\lambda = 750$~nm. Each dataset used the same 100~000 wavefront samples.

\subsection{Model hyperparameters}

\begin{table}
\begin{center}
\begin{tabular}{|c|c|c|}
    \hline
    Modulation & Zernike-basis \ac{nn}& Actuator-basis \ac{nn}  \\
    \hline
    25~mas & $10^{-9}$ & $10^{-9}$\\
    50~mas & $10^{-9}$ & $10^{-9}$\\
    75~mas & $10^{-10}$ & $10^{-9}$\\
    100~mas & $10^{-10}$ & $10^{-10}$ \\
    \hline
\end{tabular}
\caption{L1 regularisation hyperparameters for the \ac{nn}s trained in Section~\ref{sec:low-order-expts}. These values were selected using holdout validation.}
\label{tab:hyperparam-sec1}
\end{center}
\end{table}

Reconstructing in the Zernike mode basis only required small \ac{nn}s as the model only had to output 14 mode coefficients. These low-order Zernike-basis \ac{nn}s had three hidden layers with 3000, 2000 and 1000 neurons respectively and had a learning rate of $10^{-5}$. Reconstructing in the actuator mode required larger networks due to the increased complexity of the basis. These low-order actuator-basis \ac{nn}s had four hidden layers with 10~000, 7000, 5000 and 5000 neurons respectively and had a learning rate of $10^{-3}$. We applied L1 regularisation to the weights of these models and have tabulated the corresponding hyperparameter in Table.~\ref{tab:hyperparam-sec1}. These networks were trained for 2500 epochs.

The bottleneck-\ac{nn} was trained on the 25~mas modulation dataset. It had six hidden layers with 10~000, 5000, 2000, 14, 500 and 1500 neurons, respectively. We used a learning rate of $10^{-3}$, L1 regularisation hyperparameter $10^{-9}$ and trained the network for 500 epochs.  

% ------------------
\subsection{Discussion and results}
% ------------------

\begin{figure}
    \centering
    \includegraphics[width = 16.5cm]{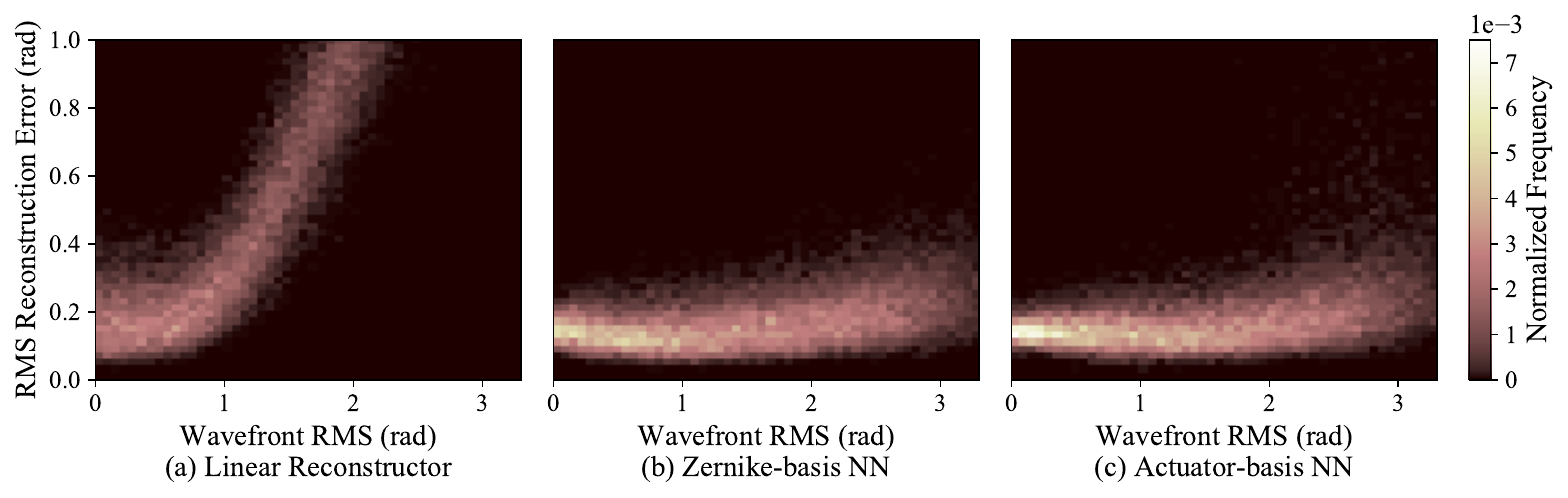}
    \caption{Heatmaps illustrating the RMS reconstruction error for different wavefront RMSs on the test data for a \ac{pywfs} modulated at 25~mas. {(a) Linear reconstructor} {(b) {low-order Zernike-basis NN}} and {(c) {low-order actuator-basis NN}}. The counts in these heatmaps are normalised by dividing the raw counts by the total number of samples.}
    \label{fig:25_mas_heatmaps}
\end{figure}

\begin{figure}
    \centering
    \subfloat[Linear scale]
        {\includegraphics[clip,height=8cm]{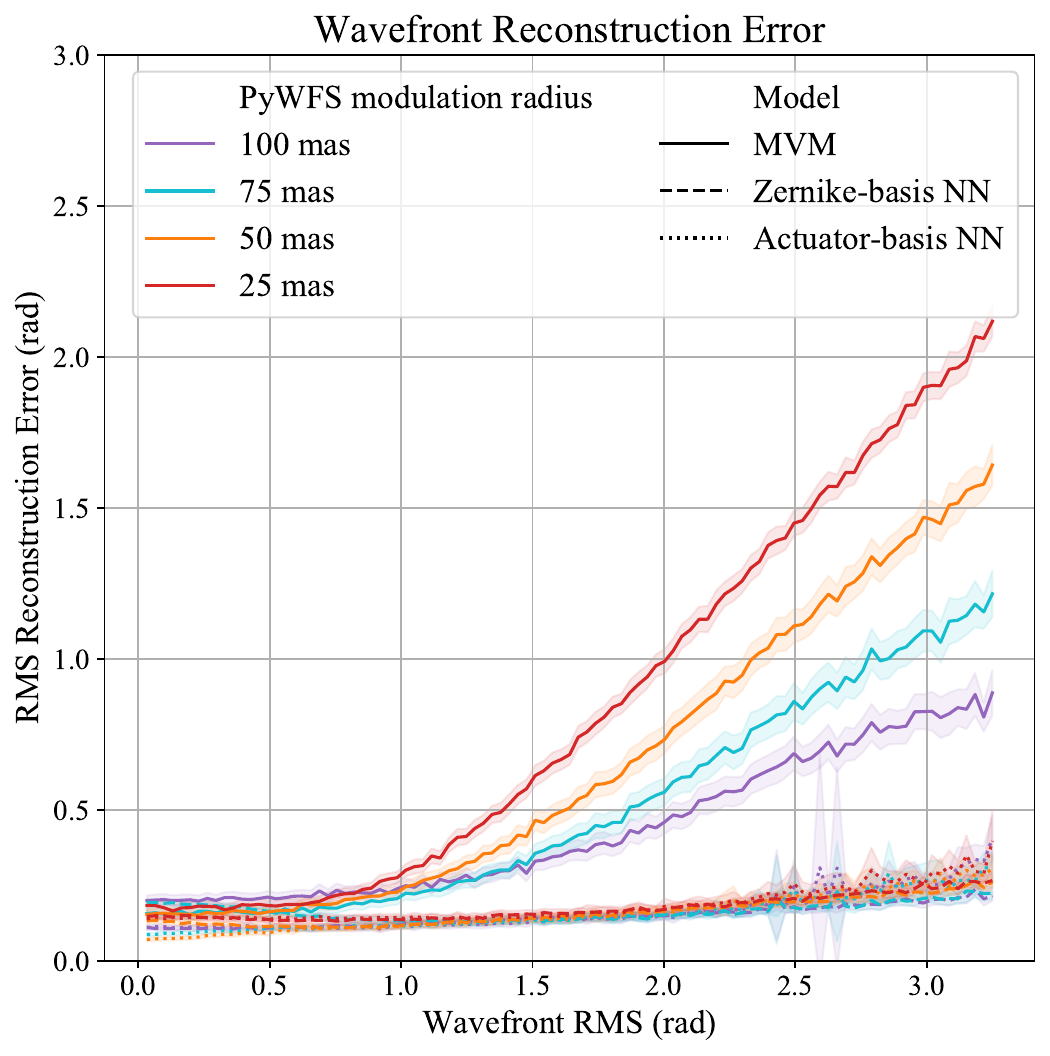}}
    \subfloat[Log scale]
        {\includegraphics[clip,height=8cm]{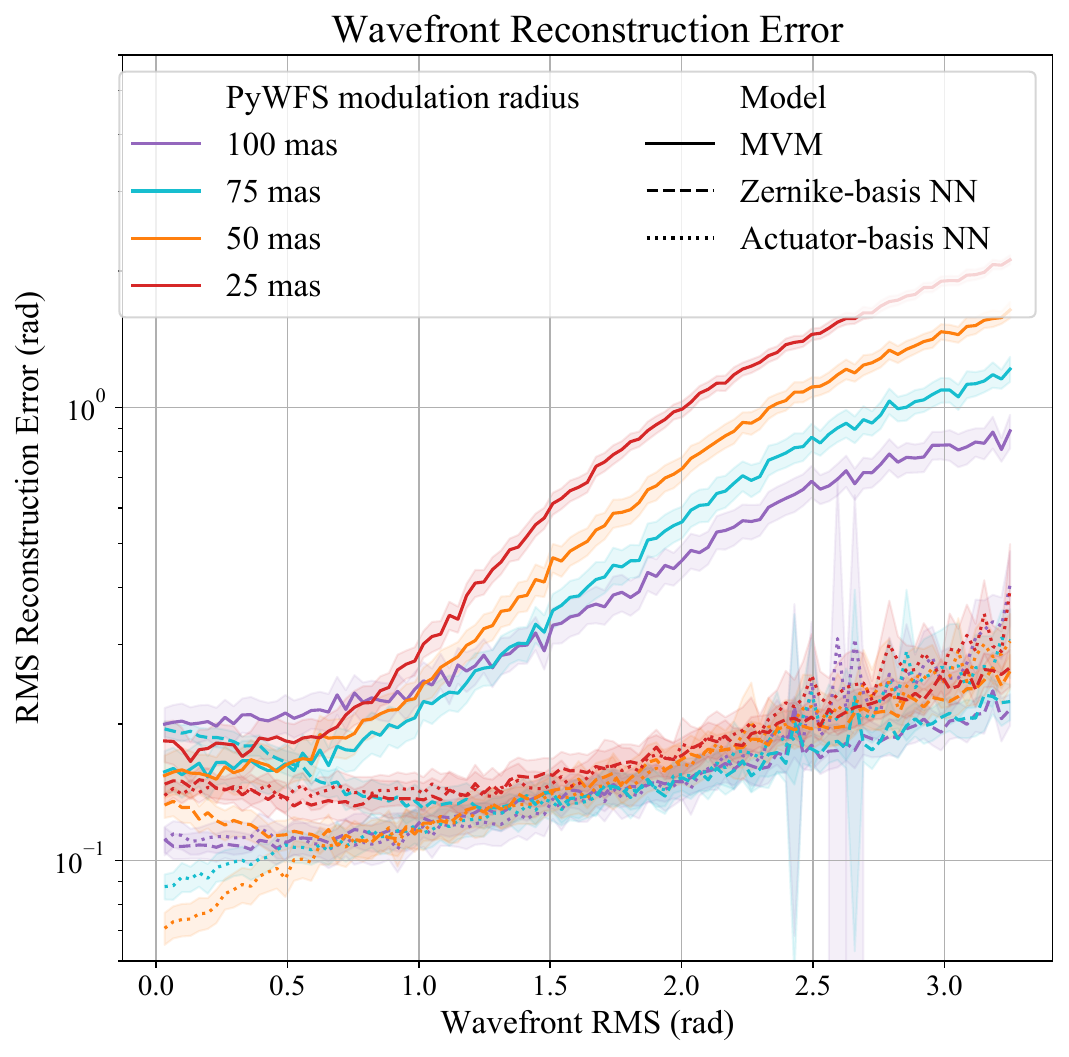}}
    \caption{Average RMS reconstruction error for the \ac{mvm} and \ac{nn}s for different pyramid modulations, for the low-order experiments. The \ac{mvm} models are plotted with a solid line, the Zernike-basis \ac{nn}s are plotted with a dashed line and the actuator-basis \ac{nn}s are plotted with a dotted line. The width of the shaded region shows 1$\sigma$ uncertainties and the colors indicate the \ac{pywfs} modulation. The \ac{mvm} only performs well at a high modulation where the \ac{pywfs} response is the most linear. However, the \ac{nn}s are seen to work successfully even at large wavefront RMSs, despite the \ac{pywfs}' nonlinearities. All \ac{nn}s have very similar performance, which implies that the actuator basis is as effective as the optimal Zernike mode basis.}
    \label{fig:zernike_reconstruction}
\end{figure}

Results showing the reconstruction accuracy for the 25~mas pyramid-modulation dataset are shown in Figure~\ref{fig:25_mas_heatmaps}. These heatmaps visually show the RMS reconstruction error against wavefront across the 20~000 test samples. Here we define RMS reconstruction error as 
\begin{equation}
    \mathlarger{\sigma}_{\begin{subarray}{l}
    \mathrm{reconstr.}\\\mathrm{error} \end{subarray}}
    = \sqrt{\frac{1}{N}\sum_{i=0}^{N} ({pred}_i - {true}_i)^2}
\end{equation}
where $pred_i$ and $true_i$ are the wavefront value (in radians) for the $i$th actuator number, for a total of $N$ active (i.e. unobstructed) actuators.

Figure~\ref{fig:zernike_reconstruction} shows a summarised view of these reconstruction errors alongside those for the 50, 75 and 100 mas modulation datasets. In Figure~\ref{fig:zernike_reconstruction}, the point at which the \ac{pywfs} response becomes nonlinear -- and the reconstruction accuracy for linear models rapidly decreases -- is clearly visible as the `knee' of the plot. 
As expected, the onset of nonlinearity occurs at a larger wavefront RMS as the modulation radius is increased. 

An important result that can be observed from Figure~\ref{fig:zernike_reconstruction} is that the \ac{nn}s are able to handle the \ac{pywfs} nonlinearities induced by introducing a wavefront whose gradient reaches into a saturated regime. Further adding to this result is that the \ac{nn} offers better performance over the \ac{mvm} at all modulations and all wavefront errors. This implies the \ac{nn} may be suitable for wavefront reconstruction from a pyramid with no modulation and, therefore maximum sensitivity. We considered this to be beyond the scope of this paper as the operation of \ac{scexao} at 0 \ac{pywfs} modulation is technically more involved as it is not typically used. Additionally, at the time the dataset was obtained, the tip-tilt drift was too large to allow for consistent unmodulated dataset, but we hope that it will be the subject of future work.

Figure.~\ref{fig:zernike_reconstruction} shows a clear offset in the RMS reconstruction errors, with the best predictions plateauing at $\sim$0.1~radians RMS reconstruction error, even as the applied wavefront RMS approaches zero. This is consistent across all model types. Inspection of the \ac{mvm} model trained on the 25 mas dataset showed that the average reconstruction residual map for samples with wavefront RMS $<$ 0.5 radians was found to have an overall wavefront RMS of $4.5\times10^{-2}$\,rad, with a maximum standard deviation for a given pixel to be 0.125\,rad, which is relatively small and indicates high confidence in our mean residual maps. We obtained similar results when inspecting the other datasets and implies that there isn’t a systematic bias in the reconstructions and that the offset is consistent with measurement noise.

It should also be noted that there is a small difference in offset between the \ac{mvm} models and the \ac{nn}s. While it is expected that both models should offer the same performance a low wavefront RMS where the \ac{pywfs} response is linear, we must remember that the \ac{mvm} has been globally optimised to perform well at all wavefront RMS. It is likely that the \ac{mvm} sacrifices some performance at low wavefront RMS to achieve better performance at high wavefront RMS.

\begin{figure}
    \centering
    \includegraphics[width=10cm]{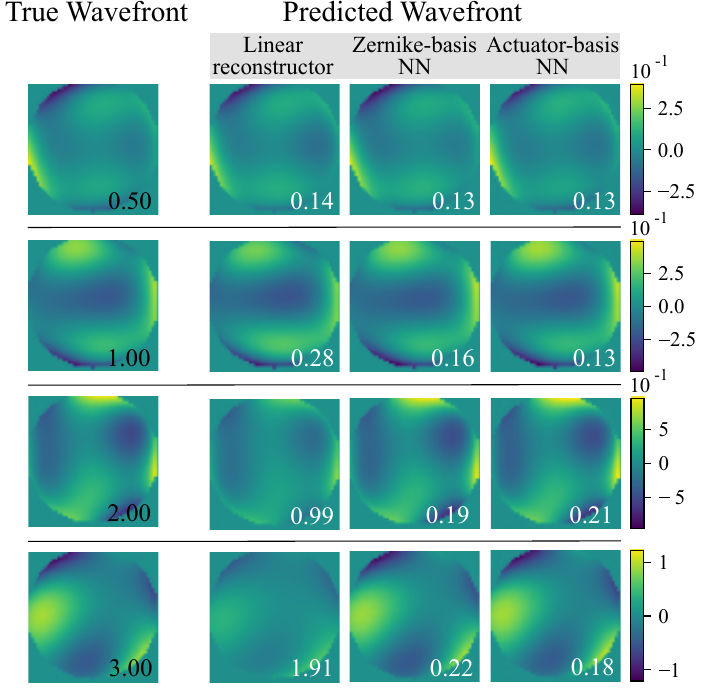}
    \caption{Sample reconstructions from the \ac{mvm}, the Zernike-basis \ac{nn} and the actuator-basis \ac{nn} on the 25~mas modulation dataset with the RMS reconstruction error in radians shown in white. The first column shows the true wavefronts that were applied to the \ac{dm} with the corresponding RMS amplitude in radians shown in black. Note the differing color scales for each row. The color scale is in units of microns.} 
    \label{fig:zernike_samples}
\end{figure}

A selection of representative wavefront samples are shown in Figure.~\ref{fig:zernike_samples}, and shows reconstructions for wavefronts with various RMSs. 

Another important outcome of this demonstration is that the actuator-basis \ac{nn}s performed equally as well as the Zernike-basis \ac{nn}s. As the Zernike-basis \ac{nn} was, by design, optimal on the Zernike dataset, this result implies that the \ac{nn} models are agnostic to the mode basis. This means an actuator basis should be suitable for reconstructing arbitrarily shaped wavefronts and additional efforts into choosing an optimal mode basis may not be necessary.

% ------------------
\subsection{Bottleneck network}
% ------------------
\label{sec:bottleneck}

\begin{figure}
	\centering
	\includegraphics[width = 17cm]{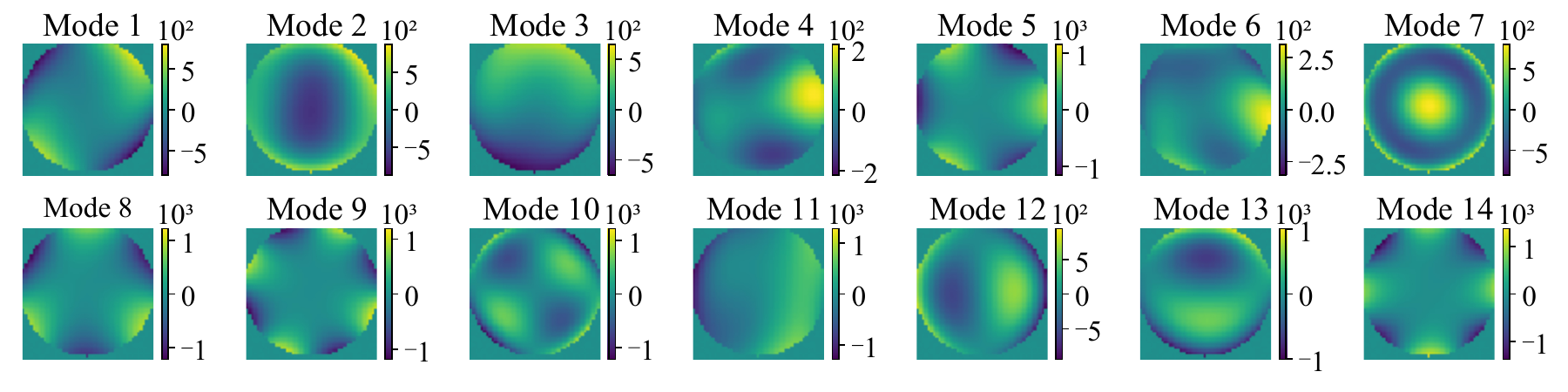}
	\caption{The mode `basis' recovered from the bottleneck-\ac{nn}, which was trained on the 25~mas modulation dataset. The color scale is in units of microns.}
	\label{fig:bottleneck}
\end{figure}

As mentioned in Section \ref{sec:NNImpl}, to gain an insight into how our \ac{nn}s `think', we produced a bottleneck network, which we trained on one of the low-order datasets used in Section~\ref{sec:low-order-expts}. This \ac{nn} performed reconstruction in the actuator basis but was designed with a central bottleneck layer with just 14 neurons, one for each mode used to generate the data. This encoder-decoder structure allowed for a mode `basis' to be recovered by breaking the network at the bottleneck and then activating each of the 14 neurons, thus emulating the two-step structure of a dual-stage modal \ac{mvm} reconstructor, from wavefront measurements into modal space and then into actuator space. The set of wavefronts represented within the bottleneck-\ac{nn} were probed by maximally activating a neuron (inputting a large value) in the bottleneck layer while setting the remaining 13 neurons to 0. For each excited neuron in the bottleneck there was a corresponding wavefront mode produced by the network. The result is shown in Figure~\ref{fig:bottleneck}. This is analogous to the modal basis of a linear system, except in this case these `modes' are created and combined in a nonlinear fashion to produce the output wavefront. Thus, these are not modes in the usual sense, but they do permit qualitative insight into the features the network is most strongly discriminating on. 

These bottleneck modes bear a striking resemblance to the Zernike modes (Figure.~\ref{fig:zernike_modes}), with difference due to the network's ability to generate and combine these modes nonlinearly. It is interesting to observe that for many modes, the \ac{nn} naturally trained to learn direct-Zernike terms rather than linear combinations thereof, which in a truly linear regime would be equivalent.
We performed this experiment a number of times, and the resultant mode basis always resembled the Zernike basis, but with slight variations to the mode shape and with random changes to the mode order.

% --------------------
\section{Reconstruction of high-order turbulence-like wavefronts}\label{sec:seeing-expts}
% --------------------
The efficacy of the NN compared to the linear method was also tested on high-order wavefront error data, with data generated via a simulated atmospheric phase screen applied to the \ac{scexao} DM. While for the linear model the standard response matrix was used, the choice of training basis for a NN is arbitrary. Any set of known input wavefronts can be used to train the model, as long as they broadly span a similar volume of wavefront space similar to that of the real, on-sky wavefronts. To that end, we simply used other randomly generated sets of Kolmogorov phase screens as the training data.

\subsection{Dataset}
The high-order data set was produced with wavefronts generated using HCIPy\cite{Por2018Hcipy} to simulate real seeing conditions.
We simulated a telescope with an 8.2\,m diameter aperture (to match the Subaru telescope) and used a wavefront sensing wavelength of 750\,nm. One dataset was produced with outer scale $L_0 = 20$\,m to simulate general atmospheric conditions, and another dataset was produced with outer scale $L_0 = 1$\,m to simulate residual seeing after initial low spatial frequency correction from the facility \ac{ao} system (AO188), after which \ac{scexao} is located. For simplicity and to ensure the absence of statistical correlation across the training set, we generated independent wavefront maps that were not temporally related. The amplitudes of these wavefront maps were then multiplied by a scaling factor to produce the desired distribution of wavefront \ac{rms}.

\begin{figure}[htp]
    \centering
    \subfloat[$L_0 = 20$\,m]
        {\includegraphics[clip,height=5cm]{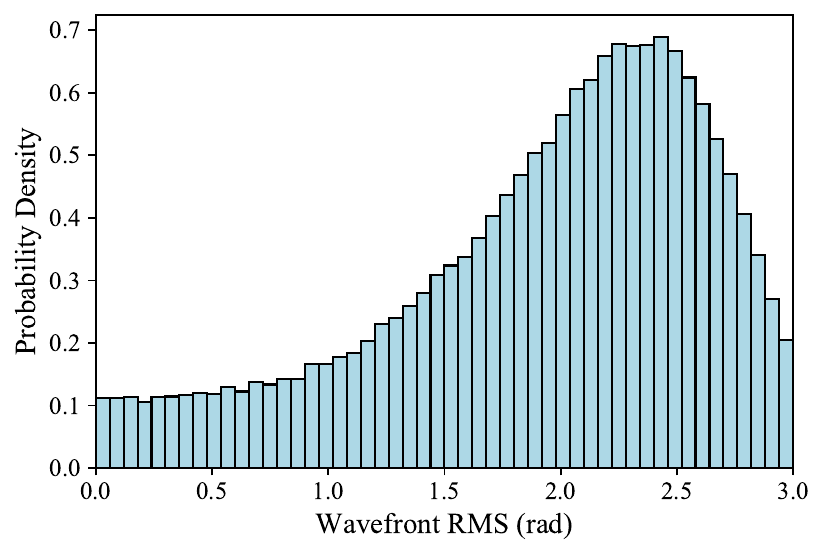}}
    \subfloat[$L_0 = 1$\,m]
        {\includegraphics[clip,height=5cm]{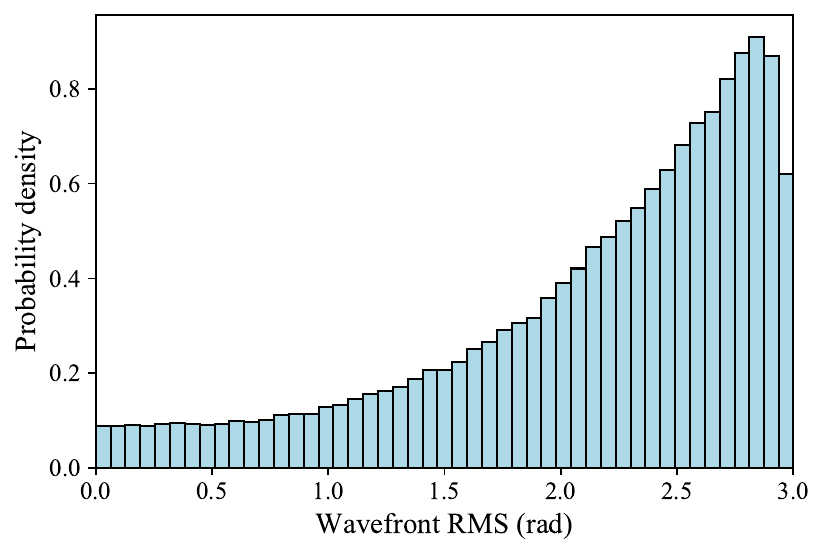}}
    \caption{Distribution of the wavefront RMSs in the training data for the two outer scales tested, normalised so that the area under the curve is 1. There are more training samples at larger wavefront RMS where the mapping from \ac{pywfs} measurements to wavefront is more complex. 
    }
    \label{fig:training_dist}
\end{figure}

The datasets were taken at a \ac{pywfs} modulation of 75~mas, which is typical for SCExAO observations. Each dataset contained 220~000 samples, which were broken into 170~000 training samples, 20~000 validation samples and 30~000 test samples. The test and validation sets were constructed to contain a uniform distribution of RMS wavefronts between 0 and 3 radians, which was achieved by multiplying wavefronts by appropriate constants. In a similar fashion, the training set was constructed to follow a cubic distribution so that there were more training samples available for wavefronts with large aberrations. This was based on an educated guess as it is expected that more training samples were required for large wavefront RMS, and the optimal distribution was not know to us. When the full $50 \times 50$ wavefront images are analysed the wavefront RMS follows a true cubic distribution, however we re-examined these distributions after masking out the inactive pixels and retaining only the active pixels (corresponding to actuators in the pupil), and results in the adjusted cubic distributions shown in (Figure.~\ref{fig:training_dist}).

\subsection{Model hyperparameters}

The network trained on the $L_0 = 20$\,m dataset had three hidden layers with 5000, 4000 and 3000 neurons and used L1 regularisation with hyperparameter $\lambda_{L1} = 10^{-9}$. It was trained for 2500 epochs with a learning rate of $10^{-4}$. The network trained on the $L_0 = 1$\,m dataset had four hidden layers with 3000, 2000, 2000 and 2000 neurons. This network was trained for 2500 epochs with a learning rate of $10^{-3}$ and found that it performed best with no regularisation. Batch normalisation was used for both models. These networks were trained using the raw, un-normalised \ac{pywfs} images. 

We chose \ac{nn}s that were overly large and complex for the problem at hand so that regularisation could be used to automatically reduce model complexity for us and select a model of appropriate complexity. It was important that we didn't start with a model that was too simple, because a simpler model cannot become more complex, but a more complex model can become simpler through regularisation.

For the linear reconstructor benchmark, the control matrix was produced as per Section ~\ref{sec:linear_reconstructors}.
The modes are roughly sorted such that as the mode number increases, so does the spatial frequency. The first 14 modes are shown in Figure.~\ref{fig:mvm_modes}. To prevent overfitting, the \ac{mvm} was regularised by selecting the number of modes kept for reconstruction which was chosen using holdout validation. This has the effect of reducing the complexity of the \ac{mvm} model and while it may perform worse on the training data, it should generalise better to unseen data. In this process, the reconstruction accuracy was measured for the same data but containing a varying number of modes. The number of modes which produced the most accurate reconstructions was then used for the subsequent experiments. For the $L_0 = 20$\,m dataset this kept the first 136 modes and for the $L_0 = 1$\,m dataset it kept the first 379 modes.

% -------------------
\subsection{Discussion and results}
% -------------------

\begin{table}
\begin{center}
\begin{tabular}{|c|c|c|c|c|}
    \cline{3-5}
    \multicolumn{2}{c|}{}& \multicolumn{3}{c|}{Wavefront RMS (rad)} \\
    \cline{3-5} \multicolumn{2}{c|}{} & 0-1 & 1-2 & 2-3 \\
    \hline
    Model & \ac{mvm} & $0.40$ & $0.73$ & $1.32$\\
    &\ac{nn} & $0.18$ & $0.31$ & $0.48$\\
    \hline
\end{tabular}
\caption{A summary of the mean RMS reconstruction error (rad) for different wavefront RMSs for the $L_0 = 20$\,m dataset comparing the \ac{mvm} (linear reconstructor) with the \ac{nn}.}
\label{tab:seeing_results_L20}
\end{center}
\end{table}

\begin{figure}
    \centering
    \includegraphics[width = 14cm]{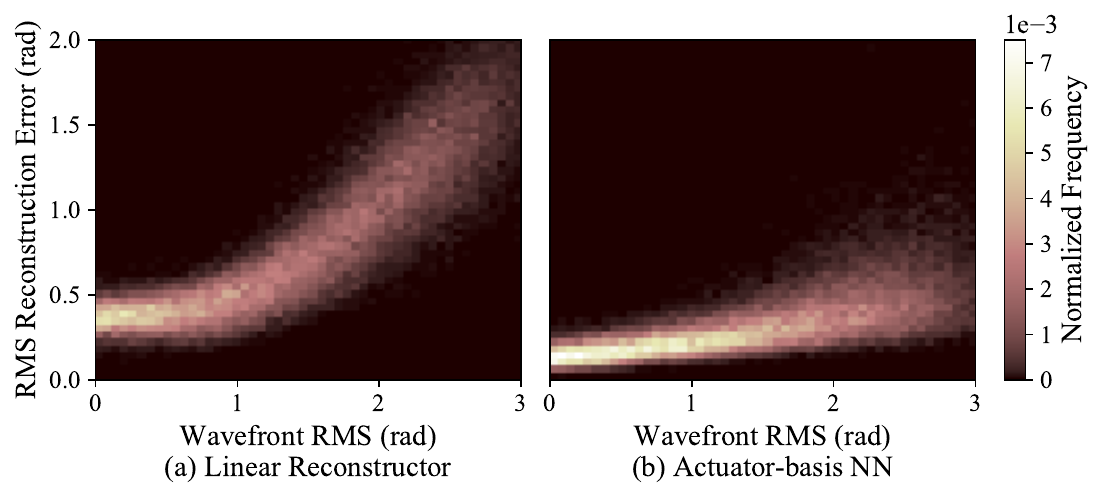}
    \caption{Heatmaps showing the mean RMS reconstruction error for samples of different wavefront RMSs, for the $L_0 = 20$\,m dataset, which simulates general seeing conditions.}
    \label{fig:l020_heatmaps}
\end{figure}

\begin{table}
\begin{center}
\begin{tabular}{|c|c|c|c|c|}
    \cline{3-5}
    \multicolumn{2}{c|}{}& \multicolumn{3}{c|}{Wavefront RMS (rad)} \\
    \cline{3-5} \multicolumn{2}{c|}{} & 0-1 & 1-2 & 2-3 \\
    \hline
    Model & \ac{mvm} & $0.52$ & $1.17$ & $2.15$\\
    &\ac{nn} & $0.25$ & $0.71$ & $1.36$\\
    \hline
\end{tabular}
\caption{A summary of the RMS reconstruction error (rad) for different wavefront RMSs for the $L_0 = 1$\,m dataset comparing the \ac{mvm} (linear reconstructor) with the \ac{nn}.}
\label{tab:seeing_results_L1}
\end{center}
\end{table}

\begin{figure}
    \centering
    \includegraphics[width = 14cm]{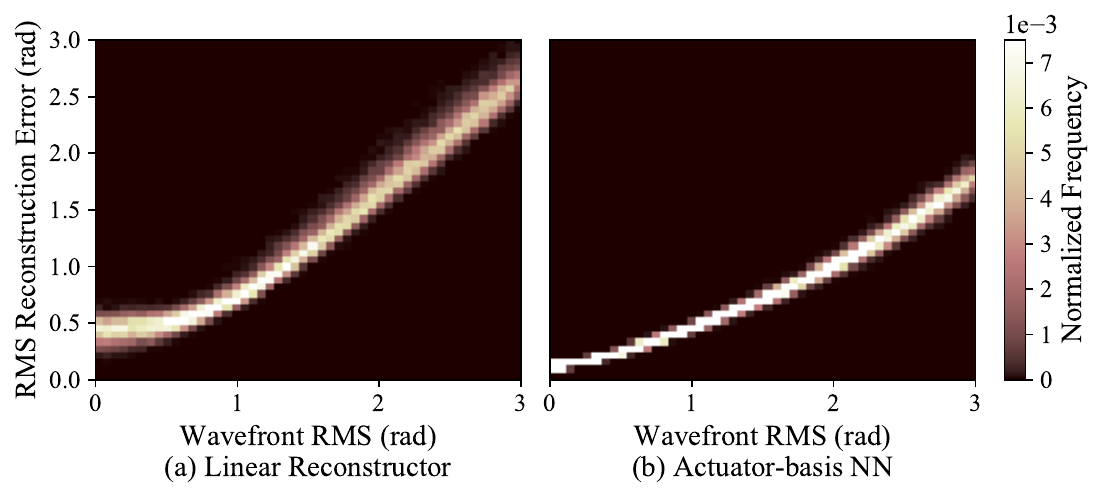}
    \caption{ Heatmaps showing the RMS reconstruction error for samples of different wavefront RMSs, for the $\mathbf{L_0 = 1} $\,m dataset, which simulates wavefronts already corrected by a low-order AO system.
    }
    \label{fig:l01_heatmaps}
\end{figure}

Heatmaps showing the \ac{rms} reconstruction error versus the wavefront \ac{rms} are illustrated in Figures~\ref{fig:l020_heatmaps} and \ref{fig:l01_heatmaps}, and tabulated summaries are provided in Tables~\ref{tab:seeing_results_L20} and \ref{tab:seeing_results_L1}. The results clearly show that the \ac{nn} outperforms the \ac{mvm} method at all wavefront \ac{rms} for both datasets. The performance improvement of the \ac{nn} over the \ac{mvm} is much more pronounced for the $L_0 = 20$\,m dataset than for the $L_0 = 1$\,m dataset. This may be because the \ac{mvm} algorithm is better optimised for the $L_0 = 1$\,m dataset (i.e. for the wavefront properties seen following correction by the facility low-order AO system), whereas the \ac{nn}s have been equally optimised for both datasets.

An obvious observation is that in both datasets the \ac{nn} outperforms the \ac{mvm} even when the wavefront \ac{rms} is small, i.e., when the \ac{pywfs} is operating in the linear regime. Initially, one would expect that both models should offer similar performance here, however it is important to recognise that the regularisation for the \ac{mvm} was optimised on the \textit{entire} dataset and most likely some performance on low wavefront errors was sacrificed in order to obtain a better performance on higher wavefront errors.

Additionally, reconstruction errors for the $L_0 = 20$\,m dataset are smaller than the $L_0 = 1$\,m datasets for both the \ac{mvm} and \ac{nn}. This indicates that the $L_0 = 1$\,m dataset is more difficult to model than the $L_0 = 20$\,m dataset owing to the higher spatial frequencies in the dataset. This is supported by the fact that the \ac{mvm} required more modes to reconstruct the $L_0 = 1$\,m than the $L_0 = 20$\,m wavefronts. Additionally, the \ac{nn} did not benefit from regularisation, which implies that it may not have been complex enough to model the trends in the data.

One potential reason that the \ac{nn} performs better on the $L_0 = 20$\,m dataset rather than the $L_0 = 1$\,m dataset is that they both have datasets of the same size but each wavefront in the $L_0 = 1$\,m dataset contains significantly more information and is harder for the model to learn. More training data is expected to improve the performance of the model, so the results presented do not necessarily demonstrate the best performance a \ac{nn} could theoretically achieve. 

\begin{figure}[htp]
    \centering
    \subfloat[$L_0 = 20$\,m]
        {\includegraphics[clip,height=8cm]{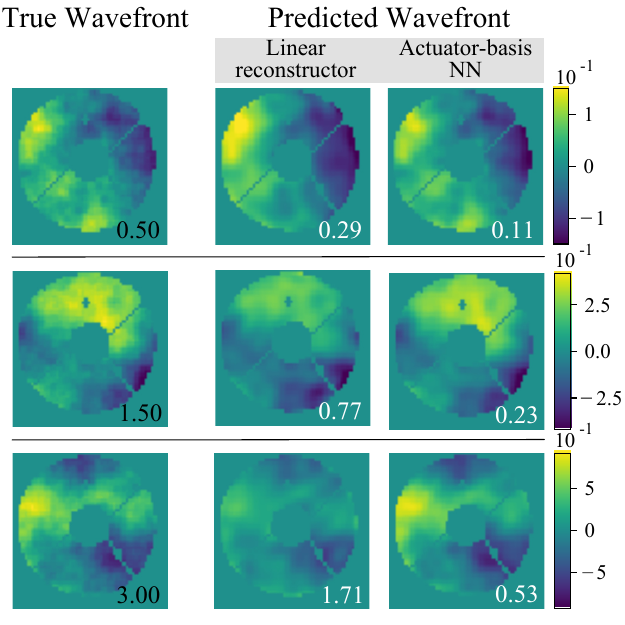}}
    \subfloat[$L_0 = 1$\,m]
        {\includegraphics[clip,height=8cm]{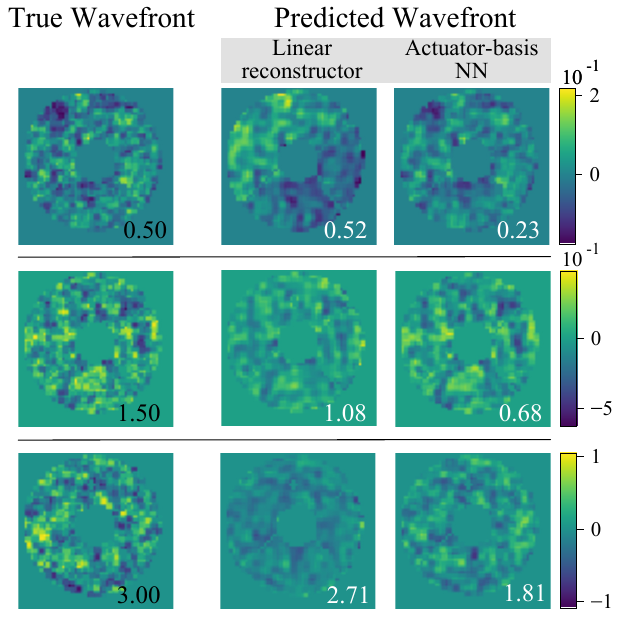}}
    \caption{Sample reconstructions from the \ac{mvm} and \ac{nn}s models with the RMS reconstruction error in radians shown in white. The first column of each subfigure shows the true wavefronts that were applied to the \ac{dm} with the corresponding wavefront RMS amplitude in radians shown in white. For display purposes, these figures were cropped to show only the active pixels. The color scale is in units of microns.
    We also see a similar effect to that in Figure.~\ref{fig:zernike_samples} where the \ac{mvm} method underestimates the wavefront amplitude as the aberrations increase.}
    \label{fig:seeing_samples}
\end{figure}

To further understand the performance of the \ac{mvm} and \ac{nn} models, we show some representative reconstruction samples in Figure.~\ref{fig:seeing_samples}. Like in Figure.~\ref{fig:zernike_samples}, we see that while the \ac{mvm} does well to reconstruct the wavefront profile, it tends to underestimate the magnitude of the wavefront, particularly when the wavefront errors become large. The \ac{nn} does not appear to suffer the same problem. Additionally, the \ac{nn} does appear to more accurately reconstruct the wavefront profile than the \ac{mvm}, most clearly seen in the $L_0=1$\,m examples in Figure \ref{fig:seeing_samples}. 

\begin{figure}[htp]
    \centering
    \subfloat[$L_0 = 20$\,m]
        {\includegraphics[clip,height=4cm]{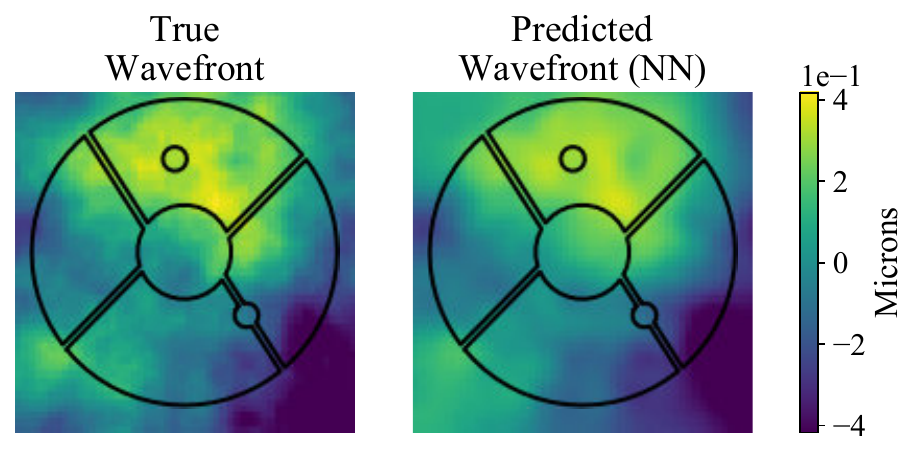}}
    \subfloat[$L_0 = 1$\,m]
        {\includegraphics[clip,height=4cm]{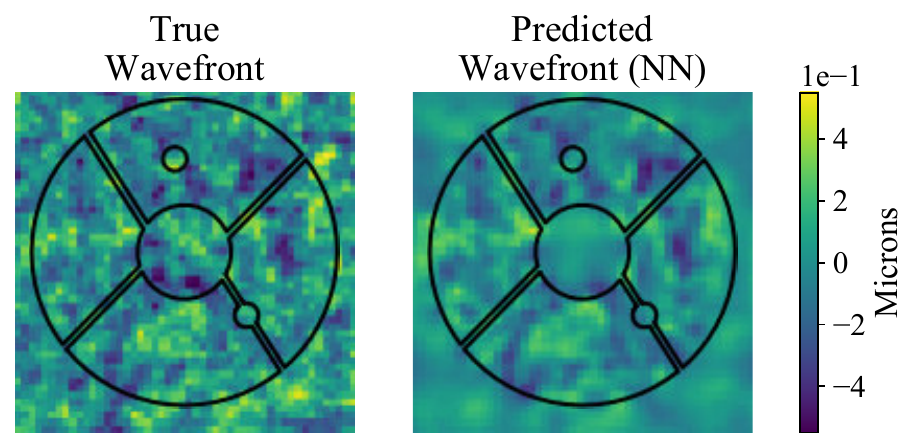}}
    \caption{Typical samples from the $L_0 = 20$\,m and $L_0 = 1$\,m datasets. \textbf{Left: True wavefront.} The 50 $\times$ 50 array of actuator commands to apply a wavefront to the \ac{dm}. \textbf{Right: \ac{nn} reconstruction.} The \ac{nn}s produce a $50 \times 50$ pixel map. Superimposed over these images is the \ac{scexao} pupil. It is interesting to note that the \ac{pywfs} is blind to anything outside of the pupil, however the \ac{nn} can makes a sensible estimate of the wavefront in these regions based on the statistics learned from the training dataset.}
    \label{fig:seeing_outfill}
\end{figure}

Figure.~\ref{fig:seeing_outfill} shows sample \ac{nn} reconstructions without the masking of inactive actuators. Superimposed over the top is an outline of the \ac{scexao} pupil. It is interesting to note that the \ac{nn} somewhat sensibly extrapolates the wavefront outside of the pupil, a region where it has no information about the wavefront. In order for the network to inpaint outside the pupil, it has presumably learnt about the statistics behind the seeing, which it then uses to extrapolate the missing wavefront regions. That is, in the cases of spatial frequencies of a similar spatial extent to the missing regions, the set of wavefront solutions that maintain spatial continuity and obey the learned distribution is well constrained. 

\section{Conclusion}
Here we have presented a number of experiments which compare the performance of the traditional linear, \ac{mvm}-based method of wavefront reconstruction with a dense \ac{nn} reconstructor. Test data was acquired using the \ac{scexao} system's \ac{pywfs}, for a range of test wavefronts applied to the system's \ac{dm}. A key finding is that the \ac{nn} was much better able to handle nonlinearities that arose from the \ac{pywfs} in the presence of large wavefront aberrations. The NN performed well on all \ac{pywfs} modulation amplitudes. while the performance of the \ac{mvm} control matrix quickly decreased as the modulation decreased. It is possible then that a \ac{nn} could be used in conjunction with an unmodulated \ac{pywfs}, thus taking full advantage of its high sensitivity. 

The \ac{mvm} method and the \ac{nn} were also compared using simulated seeing data, with and without simulated low-order pre-correction. It was found that the \ac{nn} method would consistently perform more accurate wavefront reconstructions than the \ac{mvm} method at all wavefront RMSs, suggesting it may offer improved performance under poor seeing conditions. 

While the results presented in this paper largely address an open loop \ac{ao} system, where the \ac{wfs} errors are large and more likely to push the \ac{pywfs} into the linear regime, these results still have important implications to a closed loop system. Firstly, nonlinear reconstruction at large wavefront errors makes it easier to close the \ac{ao} loop during bad conditions. While in theory you can close the loop with any large wavefront error, convergence can be slow. Thus, it can be impractical (or impossible) to close the loop during bad conditions when seeing is varying extremely rapidly, but this can be alleviated by being able to handle large wavefront errors. Secondly, in closed loop operation, there are many modes that cannot be sensed by the \ac{pywfs}. These interact nonlinearly with the lower order modes that the \ac{pywfs} can `see', which means that nonlinear reconstruction is still beneficial. 

A clear next step is to test a \ac{nn} reconstruction in closed loop, on sky. One requirement is for the \ac{nn} to reconstruct on sub-millisecond timescales. Current benchmarks demonstrate latencies well below 1 millisecond on far more complicated \ac{nn}s with 100s of layers\cite{nvidia2022}, but work needs to be done to implement this within existing AO software stacks. Either an existing low-latency machine-learning framework could be used (such as TensorRT\cite{tensorrt2022}), or since operations for a \ac{nn} can be broken down into a series of matrix operations, each followed by an evaluation of the activation function for the neurons, current AO software could be adapted to implement specific models. An ideal test configuration would be to quickly switch between NN and MVM based algorithms while observing a given target, repeated for range of seeing conditions. 

In addition to the absolute wavefront correction performance, the robustness of a NN model over different seeing conditions and across multiple observing epochs needs to be evaluated. NN models generally take some time to train (the models in this paper taking $\sim$10 hours to train on a single consumer-grade \ac{gpu}) compared to an MVM control matrix calculation. Having one pre-trained model, but with a small additional amount of `fine-tuning' training applied on sky, should be investigated.

The fully connected networks used in this paper are a very simple architecture. Further work in improving the \ac{nn} model is required, including experimenting with different network architectures, such as a convolutional neural network, which is an ideal choice as it is designed for processing image data. The translation-invariance of these network's inputs may also make them robust against \ac{pywfs} misalignment. These also have fewer free parameters, though may take longer to train. Additionally, a time domain component can be included in the model to allow for predictive control and overcome latencies in the \ac{ao} loop. Sensor fusion is also an important area of interest, but it is difficult for the \ac{mvm} method to handle a variety of different data types from multiple sources. A \ac{nn} however, should be much more amenable to this problem.

\acknowledgments % equivalent to \section*{ACKNOWLEDGMENTS}       

We would also like to acknowledge the Gadigal people of the Eora Nation as the traditional owners and custodians of the land on which the University of Sydney is built and on which some of this work was carried out. We pay our respects to their knowledge and to their Elders past, present, and emerging.

The development of SCExAO was supported by the Japan Society for the Promotion of Science (Grant-in-Aid for Research \#23340051, \#26220704, \#23103002, \#19H00703 \& \#19H00695), the Astrobiology Center of the National Institutes of Natural Sciences, Japan, the Mt Cuba Foundation and the director’s contingency fund at Subaru Telescope. The authors wish to recognize and acknowledge the very significant cultural role and reverence that the summit of Maunakea has always had within the indigenous Hawaiian community, and are most fortunate to have the opportunity to conduct observations from this mountain.

% References
\bibliography{bibliography} % bibliography data in report.bib

\begin{thebibliography}{10}

\bibitem{ragazzoni1996}
Ragazzoni, R., ``Pupil plane wavefront sensing with an oscillating prism,'' {\em Journal of Modern Optics}~{\bf 43}(2),  289--293 (1996).

\bibitem{Platt2001HistoryAP}
Platt, B. and Shack, R.~V., ``History and principles of shack-hartmann wavefront sensing.,'' {\em Journal of refractive surgery}~{\bf 17 5},  S573--7 (2001).

\bibitem{guyon2005}
Guyon, O., ``Limits of adaptive optics for high‐contrast imaging,'' {\em The Astrophysical Journal}~{\bf 629},  592–614 (Aug 2005).

\bibitem{2004SPIE.5490.1177V}
{Verinaud}, C., {Le Louarn}, M., {Korkiakoski}, V., and {Braud}, J., ``{Simulations of extreme AO: a comparison between Shack-Hartmann and pyramid-based systems},'' in [{\em Advancements in Adaptive Optics}{\nolinebreak\hspace{0.1em}]},  {Bonaccini Calia}, D., {Ellerbroek}, B.~L., and {Ragazzoni}, R., eds., {\em Society of Photo-Optical Instrumentation Engineers (SPIE) Conference Series} {\bf 5490},  1177--1188 (Oct. 2004).

\bibitem{2000SPIE.4007..416E}
{Esposito}, S., {Feeney}, O., and {Riccardi}, A., ``{Laboratory test of a pyramid wavefront sensor},'' in [{\em Adaptive Optical Systems Technology}{\nolinebreak\hspace{0.1em}]},  {Wizinowich}, P.~L., ed., {\em Society of Photo-Optical Instrumentation Engineers (SPIE) Conference Series} {\bf 4007},  416--422 (July 2000).

\bibitem{Esposito2011}
{Esposito}, S., {Riccardi}, A., {Pinna}, E., {Puglisi}, A., {Quir{\'o}s-Pacheco}, F., {Arcidiacono}, C., {Xompero}, M., {Briguglio}, R., {Agapito}, G., {Busoni}, L., {Fini}, L., {Argomedo}, J., {Gherardi}, A., {Brusa}, G., {Miller}, D., {Guerra}, J.~C., {Stefanini}, P., and {Salinari}, P.,  [{\em {Large Binocular Telescope Adaptive Optics System: new achievements and perspectives in adaptive optics}}{\nolinebreak\hspace{0.1em}]}, vol.~8149 of {\em Society of Photo-Optical Instrumentation Engineers (SPIE) Conference Series},  814902 (2011).

\bibitem{Close2013}
{Close}, L.~M., {Males}, J.~R., {Morzinski}, K., {Kopon}, D., {Follette}, K., {Rodigas}, T.~J., {Hinz}, P., {Wu}, Y.~L., {Puglisi}, A., {Esposito}, S., {Riccardi}, A., {Pinna}, E., {Xompero}, M., {Briguglio}, R., {Uomoto}, A., and {Hare}, T., ``{Diffraction-limited Visible Light Images of Orion Trapezium Cluster with the Magellan Adaptive Secondary Adaptive Optics System (MagAO)},'' {\em The Astrophysical Journal}~{\bf 774},  94 (Sept. 2013).

\bibitem{2021SPIE11823E..02W}
{Wang}, J.~J., {Delorme}, J.-R., {Ruffio}, J.-B., {Morris}, E., {Jovanovic}, N., {Echeverri}, D., {Schofield}, T., {Pezzato}, J., {Skemer}, A., and {Mawet}, D., ``{High resolution spectroscopy of directly imaged exoplanets with KPIC},'' in [{\em Society of Photo-Optical Instrumentation Engineers (SPIE) Conference Series}{\nolinebreak\hspace{0.1em}]},  {\em Society of Photo-Optical Instrumentation Engineers (SPIE) Conference Series} {\bf 11823},  1182302 (Sept. 2021).

\bibitem{Jovanovic2015}
{Jovanovic}, N., {Martinache}, F., {Guyon}, O., {Clergeon}, C., {Singh}, G., {Kudo}, T., {Garrel}, V., {Newman}, K., {Doughty}, D., {Lozi}, J., {Males}, J., {Minowa}, Y., {Hayano}, Y., {Takato}, N., {Morino}, J., {Kuhn}, J., {Serabyn}, E., {Norris}, B., {Tuthill}, P., {Schworer}, G., {Stewart}, P., {Close}, L., {Huby}, E., {Perrin}, G., {Lacour}, S., {Gauchet}, L., {Vievard}, S., {Murakami}, N., {Oshiyama}, F., {Baba}, N., {Matsuo}, T., {Nishikawa}, J., {Tamura}, M., {Lai}, O., {Marchis}, F., {Duchene}, G., {Kotani}, T., and {Woillez}, J., ``{The Subaru Coronagraphic Extreme Adaptive Optics System: Enabling High-Contrast Imaging on Solar-System Scales},'' {\em Pub. Astron. Soc. Pac.}~{\bf 127},  890 (Sept. 2015).

\bibitem{Esposito2012}
Esposito, S., Pinna, E., Quirós-Pacheco, F., Puglisi, A.~T., Carbonaro, L., Bonaglia, M., Biliotti, V., Briguglio, R., Agapito, G., Arcidiacono, C., Busoni, L., Xompero, M., Riccardi, A., Fini, L., and Bouchez, A., ``{Wavefront sensor design for the GMT natural guide star AO system},'' in [{\em Adaptive Optics Systems III}{\nolinebreak\hspace{0.1em}]},  Ellerbroek, B.~L., Marchetti, E., and Véran, J.-P., eds.,  {\bf 8447},  603 -- 612, International Society for Optics and Photonics, SPIE (2012).

\bibitem{2018SPIE10703E..0WB}
{Bouchez}, A.~H., {Angeli}, G.~Z., {Ashby}, D.~S., {Bernier}, R., {Conan}, R., {McLeod}, B.~A., {Quir{\'o}s-Pacheco}, F., and {van Dam}, M.~A., ``{An overview and status of GMT active and adaptive optics},'' in [{\em Adaptive Optics Systems VI}{\nolinebreak\hspace{0.1em}]},  {Close}, L.~M., {Schreiber}, L., and {Schmidt}, D., eds., {\em Society of Photo-Optical Instrumentation Engineers (SPIE) Conference Series} {\bf 10703},  107030W (July 2018).

\bibitem{boyer2018}
Boyer, C., ``{Adaptive optics program at TMT},'' in [{\em Adaptive Optics Systems VI}{\nolinebreak\hspace{0.1em}]},  Close, L.~M., Schreiber, L., and Schmidt, D., eds.,  {\bf 10703},  314 -- 326, International Society for Optics and Photonics, SPIE (2018).

\bibitem{2018SPIE10703E..3VC}
{Crane}, J., {Herriot}, G., {Andersen}, D., {Atwood}, J., {Byrnes}, P., {Densmore}, A., {Dunn}, J., {Fitzsimmons}, J., {Hardy}, T., {Hoff}, B., {Jackson}, K., {Kerley}, D., {Lardi{\`e}re}, O., {Smith}, M., {Stocks}, J., {V{\'e}ran}, J.-P., {Boyer}, C., {Wang}, L., {Trancho}, G., and {Trubey}, M., ``{NFIRAOS adaptive optics for the Thirty Meter Telescope},'' in [{\em Adaptive Optics Systems VI}{\nolinebreak\hspace{0.1em}]},  {Close}, L.~M., {Schreiber}, L., and {Schmidt}, D., eds., {\em Society of Photo-Optical Instrumentation Engineers (SPIE) Conference Series} {\bf 10703},  107033V (July 2018).

\bibitem{hadi2013}
{El Hadi}, K., {Vignaux}, M., and {Fusco}, T., ``{Development of a Pyramid Wave-front Sensor},'' in [{\em Proceedings of the Third AO4ELT Conference}{\nolinebreak\hspace{0.1em}]},  {Esposito}, S. and {Fini}, L., eds.,  99 (Dec. 2013).

\bibitem{2021Msngr.182...17D}
{Davies}, R., {H{\"o}rmann}, V., {Rabien}, S., {Sturm}, E., {Alves}, J., {Cl{\'e}net}, Y., {Kotilainen}, J., {Lang-Bardl}, F., {Nicklas}, H., {Pott}, J.~U., {Tolstoy}, E., {Vulcani}, B., and {MICADO Consortium}, ``{MICADO: The Multi-Adaptive Optics Camera for Deep Observations},'' {\em The Messenger}~{\bf 182},  17--21 (Mar. 2021).

\bibitem{2021Msngr.182...22B}
{Brandl}, B., {Bettonvil}, F., {van Boekel}, R., {Glauser}, A., {Quanz}, S., {Absil}, O., {Amorim}, A., {Feldt}, M., {Glasse}, A., {G{\"u}del}, M., {Ho}, P., {Labadie}, L., {Meyer}, M., {Pantin}, E., {van Winckel}, H., and {METIS Consortium}, ``{METIS: The Mid-infrared ELT Imager and Spectrograph},'' {\em The Messenger}~{\bf 182},  22--26 (Mar. 2021).

\bibitem{Hutterer2018Nonlinear}
Hutterer, V. and Ramlau, R., ``Nonlinear wavefront reconstruction methods for pyramid sensors using landweber and landweber\&\#x2013;kaczmarz iterations,'' {\em Appl. Opt.}~{\bf 57},  8790--8804 (Oct 2018).

\bibitem{Gerchberg1972A}
Gerchberg, R.~W. and Saxton, W.~O., ``{A Practical Algorithm for the Determination of Phase from Image and Diffraction Plane Pictures},'' {\em {Optik}}~{\bf {35}}({2}),  {237--246} ({1972}).

\bibitem{Frazin2018Phase}
Frazin, R.~A., ``Efficient, nonlinear phase estimation with the nonmodulated pyramid wavefront sensor,'' {\em J. Opt. Soc. Am. A}~{\bf 35},  594--607 (Apr 2018).

\bibitem{2021A&A...650A..41D}
{Deo}, V., {Gendron}, {\'E}., {Vidal}, F., {Rozel}, M., {Sevin}, A., {Ferreira}, F., {Gratadour}, D., {Galland}, N., and {Rousset}, G., ``{A correlation-locking adaptive filtering technique for minimum variance integral control in adaptive optics},'' {\em A\&A}~{\bf 650},  A41 (June 2021).

\bibitem{2020A&A...644A...6C}
{Chambouleyron}, V., {Fauvarque}, O., {Janin-Potiron}, P., {Correia}, C., {Sauvage}, J.~F., {Schwartz}, N., {Neichel}, B., and {Fusco}, T., ``{Pyramid wavefront sensor optical gains compensation using a convolutional model},'' {\em A\&A}~{\bf 644},  A6 (Dec. 2020).

\bibitem{2017arXiv170700570G}
{Guyon}, O. and {Males}, J., ``{Adaptive Optics Predictive Control with Empirical Orthogonal Functions (EOFs)},'' {\em arXiv e-prints} ,  arXiv:1707.00570 (July 2017).

\bibitem{2021SPIE11823E..06H}
{Haffert}, S.~Y., {Males}, J.~R., {Close}, L., {Long}, J., {Schatz}, L., {van Gorkom}, K., {Hedglen}, A., {Lumbres}, J., {Rodack}, A., {Guyon}, O., {Knight}, J., {Kautz}, M., and {Pearce}, L., ``{Data-driven subspace predictive control: lab demonstration and future outlook},'' in [{\em Society of Photo-Optical Instrumentation Engineers (SPIE) Conference Series}{\nolinebreak\hspace{0.1em}]},  {\em Society of Photo-Optical Instrumentation Engineers (SPIE) Conference Series} {\bf 11823},  1182306 (Sept. 2021).

\bibitem{angel1990}
Angel, J.~R., Wizinowich, P., Lloyd-Hart, M., and Sandler, D., ``{Adaptive optics for array telescopes using neural-network techniques},'' {\em Nature}~{\bf 348}(6298),  221--224 (1990).

\bibitem{sandler1991}
Sandler, D.~G., Barrett, T.~K., Palmer, D.~A., Fugate, R.~Q., and Wild, W.~J., ``{Use of a neural network to control an adaptive optics system for an astronomical telescope},'' {\em Nature}~{\bf 351}(6324),  300--302 (1991).

\bibitem{lloydhart1992}
Lloyd-Hart, M., Wizinowich, P., McLeod, B., Wittman, D., Colucci, D., Dekany, R., McCarthy, D., Anel, J., and Sandler, D., ``First results of an on-line adaptive optics system with atmospheric wavefront sensing by an artificial neural network,'' {\em The Astrophysical Journal}~{\bf 390},  L41--L44 (04 1992).

\bibitem{osborn2014}
Osborn, J., Guzman, D., de~Cos~Juez, F.~J., Basden, A.~G., Morris, T.~J., Gendron, E., Butterley, T., Myers, R.~M., Guesalaga, A., Lasheras, F.~S., Victoria, M.~G., Rodríguez, M. L.~S., Gratadour, D., and Rousset, G., ``{First on-sky results of a neural network based tomographic reconstructor: Carmen on Canary},'' in [{\em Adaptive Optics Systems IV}{\nolinebreak\hspace{0.1em}]},  Marchetti, E., Close, L.~M., and Véran, J.-P., eds.,  {\bf 9148},  1541 -- 1546, International Society for Optics and Photonics, SPIE (2014).

\bibitem{guo2006}
Guo, H., Korablinova, N., Ren, Q., and Bille, J., ``Wavefront reconstruction with artificial neural networks,'' {\em Opt. Express}~{\bf 14},  6456--6462 (Jul 2006).

\bibitem{xu2019}
Xu, Z., Yang, P., Hu, K., Xu, B., and Li, H., ``Deep learning control model for adaptive optics systems,'' {\em Appl. Opt.}~{\bf 58},  1998--2009 (Mar 2019).

\bibitem{xu2020}
Xu, Z., Wang, S., Zhao, M., Zhao, W., Dong, L., He, X., Yang, P., and Xu, B., ``Wavefront reconstruction of a shack--hartmann sensor with insufficient lenslets based on an extreme learning machine,'' {\em Appl. Opt.}~{\bf 59},  4768--4774 (Jun 2020).

\bibitem{Nousiainen:21}
Nousiainen, J., Rajani, C., Kasper, M., and Helin, T., ``Adaptive optics control using model-based reinforcement learning,'' {\em Opt. Express}~{\bf 29},  15327--15344 (May 2021).

\bibitem{nousiainen2022}
{Nousiainen, J.}, {Rajani, C.}, {Kasper, M.}, {Helin, T.}, {Haffert, S. Y.}, {V\'erinaud, C.}, {Males, J. R.}, {Van Gorkom, K.}, {Close, L. M.}, {Long, J. D.}, {Hedglen, A. D.}, {Guyon, O.}, {Schatz, L.}, {Kautz, M.}, {Lumbres, J.}, {Rodack, A.}, {Knight, J. M.}, and {Miller, K.}, ``Toward on-sky adaptive optics control using reinforcement learning - model-based policy optimization for adaptive optics,'' {\em A\&A}~{\bf 664},  A71 (2022).

\bibitem{Pou2022Adaptive}
Pou, B., Ferreira, F., Quinones, E., Gratadour, D., and Martin, M., ``Adaptive optics control with multi-agent model-free reinforcement learning,'' {\em Opt. Express}~{\bf 30},  2991--3015 (Jan 2022).

\bibitem{swanson2018}
Swanson, R., Lamb, M., Correia, C., Sivanandam, S., and Kutulakos, K., ``{Wavefront reconstruction and prediction with convolutional neural networks},'' in [{\em Adaptive Optics Systems VI}{\nolinebreak\hspace{0.1em}]},  Close, L.~M., Schreiber, L., and Schmidt, D., eds.,  {\bf 10703},  481 -- 490, International Society for Optics and Photonics, SPIE (2018).

\bibitem{He:21}
He, Y., Liu, Z., Ning, Y., Li, J., Xu, X., and Jiang, Z., ``Deep learning wavefront sensing method for shack-hartmann sensors with sparse sub-apertures,'' {\em Opt. Express}~{\bf 29},  17669--17682 (May 2021).

\bibitem{Escobar2021Wavefront}
Escobar, D. and Vera, E., ``Wavefront sensing using deep learning for shack hartmann and pyramidal sensors,'' in [{\em 2021 IEEE CHILEAN Conference on Electrical, Electronics Engineering, Information and Communication Technologies (CHILECON)}{\nolinebreak\hspace{0.1em}]},   1--5 (2021).

\bibitem{Nishizaki:19}
Nishizaki, Y., Valdivia, M., Horisaki, R., Kitaguchi, K., Saito, M., Tanida, J., and Vera, E., ``Deep learning wavefront sensing,'' {\em Opt. Express}~{\bf 27},  240--251 (Jan 2019).

\bibitem{xin2019}
Xin, Q., Ju, G., Zhang, C., and Xu, S., ``Object-independent image-based wavefront sensing approach using phase diversity images and deep learning,'' {\em Opt. Express}~{\bf 27},  26102--26119 (Sep 2019).

\bibitem{Paine:18}
Paine, S.~W. and Fienup, J.~R., ``Machine learning for improved image-based wavefront sensing,'' {\em Opt. Lett.}~{\bf 43},  1235--1238 (Mar 2018).

\bibitem{guo2019}
Guo, H., Xu, Y., Li, Q., Du, S., He, D., Wang, Q., and Huang, Y., ``{Improved Machine Learning Approach for Wavefront Sensing},'' {\em Sensors}~{\bf 19},  3533 (aug 2019).

\bibitem{andersen2019}
Andersen, T., Owner-Petersen, M., and Enmark, A., ``Neural networks for image-based wavefront sensing for astronomy,'' {\em Opt. Lett.}~{\bf 44},  4618--4621 (Sep 2019).

\bibitem{Alvarez_Diez_2008}
Diez, C.~A., Shao, F., and Bille, J., ``Pyramid and hartmann–shack wavefront sensor with artificial neural network for adaptive optics,'' {\em Journal of Modern Optics}~{\bf 55}(4-5),  683--689 (2008).

\bibitem{landman2020}
Landman, R. and Haffert, S.~Y., ``Nonlinear wavefront reconstruction with convolutional neural networks for fourier-based wavefront sensors,'' {\em Opt. Express}~{\bf 28},  16644--16657 (May 2020).

\bibitem{pope2021}
Pope, B. J.~S., Pueyo, L., Xin, Y., and Tuthill, P.~G., ``Kernel phase and coronagraphy with automatic differentiation,'' {\em The Astrophysical Journal}~{\bf 907},  40 (jan 2021).

\bibitem{Wong2021Phase}
Wong, A., Pope, B., Desdoigts, L., Tuthill, P., Norris, B., and Betters, C., ``Phase retrieval and design with automatic differentiation: tutorial,'' {\em J. Opt. Soc. Am. B}~{\bf 38},  2465--2478 (Sep 2021).

\bibitem{Landman2022}
Landman, R., Keller, C., Por, E.~H., Haffert, S., Doelman, D., and Stockmans, T., ``{Joint optimization of wavefront sensing and reconstruction with automatic differentiation},'' in [{\em Adaptive Optics Systems VIII}{\nolinebreak\hspace{0.1em}]},  Schreiber, L., Schmidt, D., and Vernet, E., eds.,  {\bf 12185},  1218589, International Society for Optics and Photonics, SPIE (2022).

\bibitem{Landman2021Self}
Landman, R., Haffert, S., Radhakrishnan, V., and Keller, C., ``Self-optimizing adaptive optics control with reinforcement learning for high-contrast imaging,'' {\em Journal of Astronomical Telescopes, Instruments, and Systems}~{\bf 7} (08 2021).

\bibitem{2021SPIE11823E..03A}
{Ahn}, K., {Guyon}, O., {Lozi}, J., {Vievard}, S., {Deo}, V., {Skaf}, N., {Belikov}, R., {Bos}, S.~P., {Bottom}, M., {Currie}, T., {Frazin}, R., {V. Gorkom}, K., {Groff}, T.~D., {Haffert}, S.~Y., {Jovanovic}, N., {Kawahara}, H., {Kotani}, T., {Males}, J.~R., {Martinache}, F., {Mazin}, B., {Miller}, K., {Norris}, B., {Rodack}, A., and {Wong}, A., ``{SCExAO: a testbed for developing high-contrast imaging technologies for ELTs},'' in [{\em Society of Photo-Optical Instrumentation Engineers (SPIE) Conference Series}{\nolinebreak\hspace{0.1em}]},  {\em Society of Photo-Optical Instrumentation Engineers (SPIE) Conference Series} {\bf 11823},  1182303 (Sept. 2021).

\bibitem{2018SPIE10703E..1EG}
{Guyon}, O., {Sevin}, A., {Gratadour}, D., {Bernard}, J., {Ltaief}, H., {Sukkari}, D., {Cetre}, S., {Skaf}, N., {Lozi}, J., {Martinache}, F., {Clergeon}, C., {Norris}, B., {Wong}, A., and {Males}, J., ``{The compute and control for adaptive optics (CACAO) real-time control software package},'' in [{\em Adaptive Optics Systems VI}{\nolinebreak\hspace{0.1em}]},  {Close}, L.~M., {Schreiber}, L., and {Schmidt}, D., eds., {\em Society of Photo-Optical Instrumentation Engineers (SPIE) Conference Series} {\bf 10703},  107031E (July 2018).

\bibitem{Nair2010Rectified}
Nair, V. and Hinton, G.~E., ``Rectified linear units improve restricted boltzmann machines,'' in [{\em Proceedings of the 27th International Conference on International Conference on Machine Learning}{\nolinebreak\hspace{0.1em}]},  {\em ICML'10},  807–814, Omnipress, Madison, WI, USA (2010).

\bibitem{Ioffe2015Batch}
Ioffe, S. and Szegedy, C., ``Batch normalization: Accelerating deep network training by reducing internal covariate shift,'' in [{\em Proceedings of the 32nd International Conference on Machine Learning}{\nolinebreak\hspace{0.1em}]},  Bach, F. and Blei, D., eds., {\em Proceedings of Machine Learning Research} {\bf 37},  448--456, PMLR, Lille, France (07--09 Jul 2015).

\bibitem{Keras2015}
Chollet, F. et~al., ``Keras.'' \url{https://keras.io} (2015).

\bibitem{Tensorflow2015}
Abadi, M., Agarwal, A., Barham, P., Brevdo, E., Chen, Z., Citro, C., Corrado, G.~S., Davis, A., Dean, J., Devin, M., Ghemawat, S., Goodfellow, I., Harp, A., Irving, G., Isard, M., Jia, Y., Jozefowicz, R., Kaiser, L., Kudlur, M., Levenberg, J., Man\'{e}, D., Monga, R., Moore, S., Murray, D., Olah, C., Schuster, M., Shlens, J., Steiner, B., Sutskever, I., Talwar, K., Tucker, P., Vanhoucke, V., Vasudevan, V., Vi\'{e}gas, F., Vinyals, O., Warden, P., Wattenberg, M., Wicke, M., Yu, Y., and Zheng, X., ``{TensorFlow}: Large-scale machine learning on heterogeneous systems,'' (2015).
\newblock Software available from tensorflow.org.

\bibitem{Wong2021}
Wong, A.~P., Norris, B. R.~M., Tuthill, P.~G., Scalzo, R., Lozi, J., Vievard, S.~B., and Guyon, O., ``{Predictive control for adaptive optics using neural networks},'' {\em Journal of Astronomical Telescopes, Instruments, and Systems}~{\bf 7}(1),  1 -- 22 (2021).

\bibitem{Tibshirani1996Lasso}
Tibshirani, R., ``Regression shrinkage and selection via the lasso,'' {\em Journal of the Royal Statistical Society. Series B (Methodological)}~{\bf 58}(1),  267--288 (1996).

\bibitem{srivastava2014a}
Srivastava, N., Hinton, G., Krizhevsky, A., Sutskever, I., and Salakhutdinov, R., ``Dropout: A simple way to prevent neural networks from overfitting,'' {\em Journal of Machine Learning Research}~{\bf 15}(56),  1929--1958 (2014).

\bibitem{Por2018Hcipy}
Por, E.~H., Haffert, S.~Y., Radhakrishnan, V.~M., Doelman, D.~S., Van~Kooten, M., and Bos, S.~P., ``{High Contrast Imaging for Python (HCIPy): an open-source adaptive optics and coronagraph simulator},'' in [{\em Adaptive Optics Systems VI}{\nolinebreak\hspace{0.1em}]},  {\em Proc. {{SPIE}}} {\bf 10703} (2018).

\bibitem{nvidia2022}
NVIDIA, ``Data center deep learning product performance.'' \url{https://developer.nvidia.com/deep-learning-performance-training-inference} (June 2022).

\bibitem{tensorrt2022}
NVIDIA, ``Nvidia tensorrt.'' \url{https://developer.nvidia.com/tensorrt} (June 2022).

\end{thebibliography}
\bibliographystyle{spiebib} % makes bibtex use spiebib.bst

\end{document}